\documentclass{article}
\usepackage[eandd,final]{neurips_2026}
\makeatother
\usepackage{tcolorbox}
\usepackage{subcaption}
\usepackage[utf8]{inputenc}
\usepackage[T1]{fontenc}
\usepackage{microtype}
\usepackage{amsmath,amssymb,amsfonts,mathtools,amsthm}
\usepackage{booktabs}
\usepackage{array}
\usepackage{multirow}
\usepackage{graphicx}
\usepackage{xcolor}
\usepackage{comment}
\usepackage{float}
\usepackage{url}
\usepackage{bbm}
\usepackage{algorithm}
\usepackage{algpseudocode}

\usepackage{enumitem}
\usepackage{nicefrac}
\usepackage{wrapfig}
\usepackage{hyperref}
\hypersetup{
  colorlinks=false,
  pdfborder={0 0 0.8},
  citebordercolor={0 0.65 0},  
  linkbordercolor={1 0 0},     
  urlbordercolor={0 0 1}       
}

\definecolor{darkkhaki}{RGB}{189,183,107}
\definecolor{teal}{RGB}{0,128,128}
\title{BayesJudge: Uncertainty-Aware Bayesian Meta-Evaluation of Human and LLM Judgments}

\author{%
  Jiahao Zhang\textsuperscript{*}\\
  South China Normal University\\
  \texttt{jerrycu@m.scnu.edu.cn}
  \And
  Pengbin Feng\textsuperscript{*\dag}\\
  University of Southern California\\
  \texttt{pengbinf@alumni.usc.edu}
  \AND 
  Chunlei Meng\\
  Fudan University\\
  \texttt{clmeng23@m.fudan.edu.cn}
  \And
  Hang He\\
  East China Normal University\\
  \texttt{hang.he@stu.ecnu.edu.cn}
  \AND
  Tianyong Hao\textsuperscript{\dag}\\
  South China Normal University\\
  \texttt{haoty@m.scnu.edu.cn}
}
\date{}

\newtheorem{proposition}{Proposition}

\newcommand{\softmax}{\operatorname{softmax}}
\newcommand{\Cat}{\operatorname{Cat}}
\newcommand{\Dir}{\operatorname{Dir}}

\newcommand{\Var}{\operatorname{Var}}

\newcommand{\ESS}{\operatorname{ESS}}

\begin{document}
\maketitle
\begingroup
\renewcommand{\thefootnote}{*}
\footnotetext[0]{Equal contribution}
\renewcommand{\thefootnote}{\dag}
\footnotetext[0]{Corresponding authors}
\endgroup
\begin{abstract}
AI evaluation pipelines often produce conflicting judgments rather than clean labels. Yet existing methods largely sidestep this conflict by collapsing disagreement into a single hard label. As a result, item ambiguity, rater biases, and the position preferences of LLM judges are erased together: evaluation outcomes appear deterministic while concealing the underlying uncertainty, and may even systematically favor one response. This conflict is especially acute in pairwise LLM evaluation, where disagreement arises from ambiguous items, heterogeneous raters, and order-sensitive LLM judges. We propose \textbf{BayesJudge}, an online Bayesian meta-evaluation layer for conflicting human-LLM judgment streams. BayesJudge treats the latent preference of each item as a latent variable and jointly models two sources of evaluator bias: rater-specific confusion matrices for human annotation bias, and an explicit order parameter for the position preference of LLM judges, thereby separating response quality from evaluator bias. We derive an exact online posterior recursion and adopt a scalable Rao-Blackwellized assumed-density sequential Monte Carlo (SMC) approximation for streaming inference. Consequently, each comparison yields not a single hard label but a posterior distribution over each response winning (as well as a tie or ambiguity state), together with posterior estimates of the bias parameters, making evaluation uncertainty explicit and quantifiable. On synthetic data, BayesJudge accurately recovers human annotation biases and LLM position biases online, and strictly outperforms variants that freeze either bias channel in held-out predictive log-likelihood. On the real-world SummEval dataset, BayesJudge detects systematic order effects in LLM judgments, distinguishes the behavioral signatures of experts and crowdworkers without rater metadata, and yields posterior uncertainty that reflects genuine human disagreement. Our code is available at \url{https://github.com/Killerofthecard/BayesJudge.git}.
\end{abstract}
\section{Introduction}
\label{sec:intro}
Modern AI systems are increasingly trained and evaluated from pairwise judgment data rather than objective labels. In production annotation and evaluation pipelines, the same item can receive conflicting labels from different raters, or even from the same rater at different times. These conflicts often stem not merely from annotation mistakes, but also from underspecified standard operating procedures or rubrics, ambiguous items, corner cases, subjective preferences, expertise differences, fatigue, and systematic cognitive biases \citep{dawid1979maximum,whitehill2009whose,uma2021learning,abercrombie2025consistency,fleisig2023majority,fleisig2024perspectivist}. Yet mainstream practice still collapses such conflicting observations into a single majority label, discarding the diagnostic information carried by the conflict itself: we can no longer tell whether high disagreement on an item reflects its intrinsic ambiguity or the systematic biases of individual evaluators, nor can we quantify how much trust each evaluation conclusion deserves. 
To address the prohibitive cost of human annotation, LLM-as-a-judge has emerged as a scalable surrogate for pairwise comparison at scale \citep{zheng2023judging,chiang2024chatbot,lambert2025rewardbench,tan2024judgebench}. However, LLM judges exhibit systematic biases that compromise their reliability as evaluation proxies \citep{li2025evaluating, pombal2026self, haldar2025rating, park2024offsetbias, wei2024systematic}, most notably position bias, which favors responses presented in specific slots \citep{shi2025judging, zheng2023judging}. This bias is not random noise but structural distortions inherent to the judge model's training and inference dynamics, making them particularly insidious when LLM judgments are naively aggregated or used to supervise downstream model training.
\begin{figure*}[t]
  \centering
  \includegraphics[width=1\textwidth]{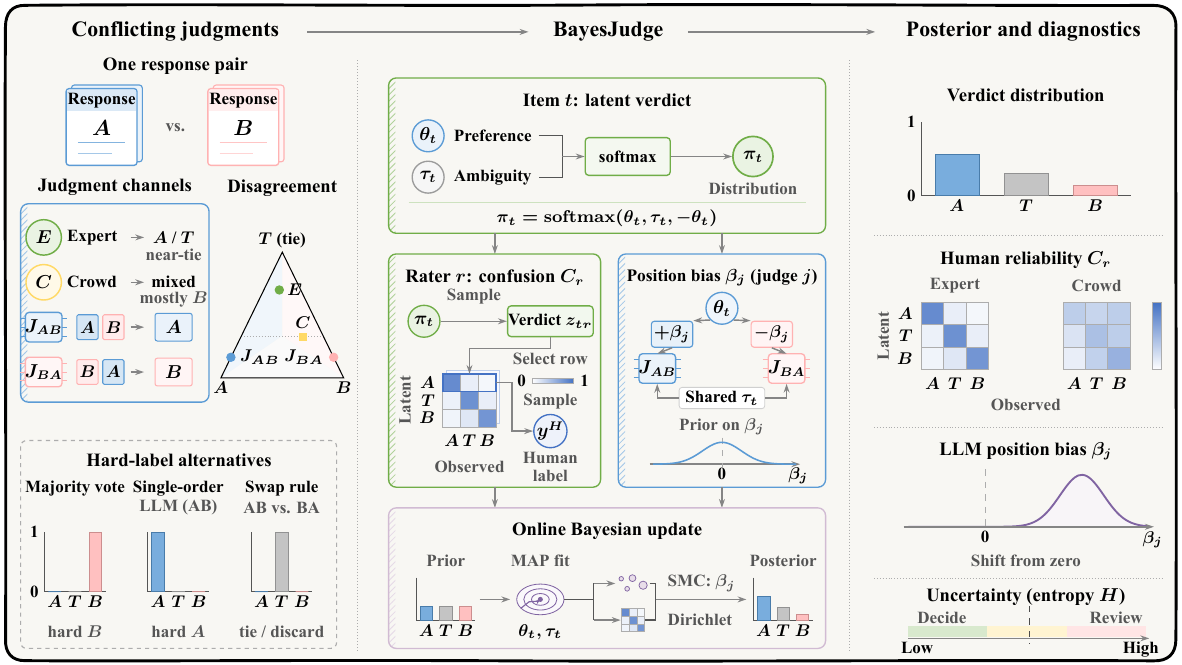}
  \caption{\textbf{BayesJudge overview.}
\textbf{Left:} For one response pair, experts ($E$) lean toward $A$ or a near tie, crowdworkers ($C$) toward $B$, and an LLM judge flips its verdict across presentation orders ($J_{AB}\!\to\!A$, $J_{BA}\!\to\!B$); the simplex locates each evaluator's judgments over $\{A,T,B\}$, and the inset shows that hard-label alternatives resolve this conflict differently.
\textbf{Center:} Preference $\theta_t$ and ambiguity $\tau_t$ define the verdict distribution $\pi_t$; a rater-specific latent verdict $z_{tr}\sim\pi_t$ selects the row of
$C_r$ generating label $y_{tr}^H$, while position bias $\beta_j$ enters with opposite signs under AB/BA orders. Online inference combines local MAP fits, SMC bias particles, and Dirichlet updates.
\textbf{Right:} a posterior verdict distribution, evaluator diagnostics, and entropy $H$ for decision or review.
$T$: tie/ambiguity; $y^H$ abbreviates $y_{tr}^H$; all plots are schematic.}
  \label{fig:motivation}
\end{figure*}


Existing work has largely proceeded along two disconnected lines. On one hand, LLM judge debiasing methods \citep{soumik2026judging,li2025calibraeval,zhao2026care} focus on mitigating systematic artifacts such as position bias, yet remain disconnected from their alignment target, namely human annotator biases. On the other hand, human annotator noise modeling \citep{gemalmaz2021accounting,li2023mitigating} estimates rater confusion structure, but overlooks how the biases of LLM judges propagate into aggregate evaluation outcomes. More critically, both paradigms ultimately produce single-point aggregate estimates that suppress genuine noise and disagreement. Such suppression is particularly harmful in open-ended pairwise evaluation without objective ground truth, where the very presence of substantive disagreement carries diagnostic value about item ambiguity and evaluator reliability, precisely the information that single-point estimates erase.

A natural question thus arises: 

\emph{When human annotations themselves exhibit disagreement and LLM evaluations carry systematic biases, how can we transform these conflicting pieces of evidence into quantifiable assessment inferences?}

In this paper, we propose \textbf{BayesJudge}, a Bayesian meta-evaluation framework for conflicting judgments, whose central tenet is to disentangle conflicting human and LLM judgments into item ambiguity, rater noise, and presentation effects, rather than average them away. BayesJudge frames evaluation as a streaming scenario: evaluation tasks arrive continuously, and a well-designed system should improve over time by accumulating evaluator bias patterns to infer stable preference structures, which is critical for reliable aggregation. Figure~\ref{fig:motivation} illustrates how BayesJudge turns conflicting judgments into posterior verdicts and evaluator diagnostics: for the same response pair, experts may favor A or perceive a near tie, crowdworkers may lean toward B, and an LLM judge may exhibit position bias. BayesJudge jointly models item-level ambiguity, human rater confusion patterns, and LLM position bias, performing principled probabilistic aggregation over biased human and LLM judgments. Specifically, the latent verdict of each comparison is modeled as a latent variable determined jointly by item preference and ambiguity, and the label of each rater or LLM judge is treated as a noisy observation of this latent variable through its own bias channel, rather than a direct readout of the truth. Under this generative model, aggregation is no longer voting or weighted averaging, but joint posterior inference over the latent verdict and evaluator parameters: for each comparison, the framework outputs a posterior verdict distribution over $\{A, \text{Tie}, B\}$ and attributes the conflict to item-level ambiguity signals, systematic rater biases, or LLM order effects. At the protocol level, the framework relies on tie-open labels and paired AB/BA order swaps as identification instruments, which allow these factors to be disentangled and quantified during streaming inference, thereby supporting online evaluation that is continually calibrated as data accumulates.


\textbf{Our contributions are as follows:} \textbf{(i)} We propose BayesJudge, a Bayesian meta-evaluation framework that decomposes conflicting human and LLM judgments into item ambiguity, rater noise, and presentation-order effects, producing posterior verdict distributions and evaluator diagnostics rather than a single hard label. \textbf{(ii)} We prove that tie-open labels and paired order swaps serve as identification instruments for item-level tie mass and LLM position bias, respectively. \textbf{(iii)} We derive an exact online posterior recursion over the global evaluator parameters, together with a scalable Rao-Blackwellized assumed-density sequential Monte Carlo (SMC) approximation for streaming inference. \textbf{(iv)} Experiments on synthetic data and SummEval show accurate recovery of rater and judge biases, unsupervised discovery of distinct expert and crowdworker signatures, detection of systematic LLM position effects, and posterior uncertainty significantly correlated with genuine human disagreement.

\section{Related Work}
\paragraph{Conflicting labels and disagreement-aware annotation.}
Classical aggregation methods treat annotations as noisy observations of a
latent label \citep{dawid1979maximum,whitehill2009whose}. Recent work views
disagreement as signal rather than noise \citep{uma2021learning}, and studies
in collective decision-making further model evaluator quality as
context-dependent: ExpertiseTrees partition the problem space into regions of
differing expertise and adapt aggregation to each region
\citep{pmlr-v202-abels23a}, while \citet{abels_ijcai} apply such localized
weighting to hybrid human--LLM crowds for bias mitigation. BayesJudge shares
the view that disagreement should be disentangled rather than averaged away,
but differs in scope and target: it assumes a stationary per-rater confusion
matrix within the evaluation domain, and infers latent pairwise verdicts
without gold labels by jointly modeling item ambiguity, rater confusion, and
presentation-order effects.
\paragraph{Pairwise LLM evaluation.} Pairwise comparison has become the dominant paradigm for LLM evaluation, underlying crowdsourced leaderboards such as Chatbot Arena \citep{chiang2024chatbot}, as well as automated benchmarks including MT-Bench \citep{zheng2023judging}, RewardBench \citep{lambert2025rewardbench}, and JudgeBench \citep{tan2024judgebench}. These platforms typically collect preference-style judgments and aggregate them into scalar win rates or Elo scores. By design, this aggregation treats conflicting or low-confidence judgments as noise to be averaged out, and it does not explicitly model why a judge might disagree with itself or with humans. 
\paragraph{LLM evaluator bias and instability.} Empirical studies have shown that LLM judges are not invariant measurement devices: their verdicts can depend on presentation order, response length, prompt phrasing, and sampling variation \citep{zheng2023judging,shi2025judging,wei2024systematic,haldar2025rating}. Among these, position bias is the most widely studied artifact, often detected via ad-hoc swap tests that compare AB and BA outputs.
\paragraph{Calibration and online inference.} Recent methods calibrate win rates or train distributional judges \citep{gao2024bayesian,chen2025distributional,polo2025bridging}. We instead model the observation process itself, using standard online Bayes filtering adapted to the local--global structure of pairwise evaluation.

\section{BayesJudge: Model and Online Inference}
\label{sec:model}

\begin{table}[t]
  \caption{Notation.}
  \label{tab:notation}
  \centering
  \small
  \begin{tabular}{ll}
    \toprule
    Symbol & Meaning \\
    \midrule
    \multicolumn{2}{l}{\emph{Data}} \\
    ${D}_t=(x_t, Y_t^H, Y_t^L)$ & one comparison: context, human and LLM labels \\
    ${R}_t$, ${J}_t$ & human raters / LLM judges of item $t$ \\
    $\mathcal{K}=\{A,T,B\}$ & verdict space ($T$: tie) \\
    \addlinespace
    \multicolumn{2}{l}{\emph{Item-level latent variables}} \\
    $\eta_t=(\theta_t,\tau_t)$ & preference strength; tie/ambiguity mass \\
    $\pi_t=g(\eta_t)$, $z_{tr}$ & verdict distribution; latent verdict \\
    \addlinespace
    \multicolumn{2}{l}{\emph{Evaluator parameters}} \\
    $C_r$ & confusion matrix of rater $r$ \\
    $\beta_j$ & position bias of judge $j$ \\
    \bottomrule
  \end{tabular}
\end{table}
We formalize the streaming pairwise evaluation setting, the observation model, and the inference procedure. Table~\ref{tab:notation} summarizes the notation.

\paragraph{Data and coordinates.}
We observe a stream of pairwise comparisons $D_{1:N}=\{D_t\}_{t=1}^N$, where each item is $D_t=(x_t,Y_t^H,Y_t^L)$ with context $x_t=(q_t,r_t^A,r_t^B)$: $q_t$ is the query or source document and $r_t^A,r_t^B$ are the two candidate responses. Human labels are $Y_t^H=\{y_{tr}^H\}_{r\in R_t}$, where $R_t$ denotes the set of human raters observed for item $t$, and LLM labels are paired AB/BA judgments $Y_t^L=\{y_{tj}^{L,(AB)},y_{tj}^{L,(BA)}\}_{j\in J_t}$ over $\mathcal K=\{A,T,B\}$, where $J_t$ denotes the set of LLM judges queried for item $t$ and $T$ denotes tie, ambiguity, or effective equivalence.
Let $\tilde y_{tj}^{L,(o)}$ denote the raw prompt-position output under order $o\in\{AB,BA\}$, where the prompt emits $A$, $B$, or $C$ and $C$ denotes a tie. We use
\begin{equation}
  h_{AB}(A)=A,\qquad h_{AB}(B)=B,\qquad h_{AB}(C)=T,
  \label{eq:hAB}
\end{equation}
for the AB prompt. For the BA prompt, Assistant A is the original response $B$ and Assistant B is the original response $A$, hence
\begin{equation}
  h_{BA}(A)=B,\qquad h_{BA}(B)=A,\qquad h_{BA}(C)=T.
  \label{eq:hBA}
\end{equation}
We set $y_{tj}^{L,(o)}=h_o(\tilde y_{tj}^{L,(o)})$ and define $Y_t^L=\{y_{tj}^{L,(AB)},y_{tj}^{L,(BA)}\}_{j\in J_t}$. Thus the coordinates $A$ and $B$ in every formula always refer to the original candidates, not to prompt positions.

\paragraph{Latent verdict.}
Each item has local latent parameters $\eta_t=(\theta_t,\tau_t)\in\mathbb R^2$. The parameter $\theta_t$ controls directional preference and $\tau_t$ controls tie or ambiguity mass. We use the Bradley--Terry--Davidson-style map~\citep{bradley1952rank, davidson1970ties} with prior $\eta_t\sim\mathcal N(\mu_0,\Sigma_0)$:
\begin{equation}
  \pi_t=g(\eta_t)=\softmax(\theta_t,\tau_t,-\theta_t)
  =\left(\frac{e^{\theta_t}}{Z_t},\frac{e^{\tau_t}}{Z_t},\frac{e^{-\theta_t}}{Z_t}\right),
  \qquad
  Z_t=e^{\theta_t}+e^{\tau_t}+e^{-\theta_t}.
  \label{eq:pi}
\end{equation}
For the human channel, each rater's label is generated from its own latent verdict $z_{tr}\mid\eta_t\sim\mathrm{Cat}(g(\eta_t))$. Labels of different raters on the same item are coupled only through the shared verdict distribution $\pi_t=g(\eta_t)$, in the spirit of Dawid--Skene-style confusion models.
\paragraph{Human and LLM observation channels.}
Each human rater $r$ has a row-stochastic confusion matrix $C_r\in[0,1]^{3\times 3}$, with rows indexing the item-level latent verdict and columns indexing observed labels. Conditional on $z_{tr}=k$,
\begin{equation}
  p(y_{tr}^H=\ell\mid z_{tr}=k,C_r)=C_r[k,\ell],
  \qquad
  C_r[k,:]\sim\Dir(\lambda_{\rm off}\mathbf 1+(\lambda_{\rm diag}-\lambda_{\rm off})e_k),
  \label{eq:human_channel}
\end{equation}
where $\lambda_{\rm diag}>0$ and $\lambda_{\rm off}>0$ . The additional inequality $\lambda_{\rm diag}>\lambda_{\rm off}$ encodes a diagonal-dominance assumption: a rater is more likely to report the latent verdict than any particular alternative, reflecting the prior domain knowledge that annotators who have passed training or qualification screening are generally correct. A sensitivity analysis of this assumption is provided in Appendix~\ref{app:init_analysis}. 
We further assume a stationary rater model: conditioning is on the latent verdict only, $p(y_{tr}^H=\ell\mid z_{tr}=k, r, x_t)=C_r[k,\ell]$, hence the confusion mechanism does not vary with the item context $x_t$; item dependence enters solely through the item-specific verdict distribution $\pi_t$. After marginalizing $z_{tr}$, the human-label likelihood for item $t$ factorizes across raters as
\begin{equation}
  H_t(\eta_t,C) := p(Y_t^H\mid \eta_t,C)
  = \prod_{r\in R_t} \sum_{k\in\mathcal K} g_k(\eta_t)\, C_r[k,y_{tr}^H]
  = \prod_{r\in R_t} \big( C_r^\top g(\eta_t) \big)_{y_{tr}^H},
  \label{eq:human_joint_likelihood}
\end{equation}
where $C=\{C_r\}_{r\in\mathcal R}$. Each factor is the per-rater marginal predictive. 

For an LLM judge $j$, let $\beta_j\in\mathbb R$ denote presentation-position bias and let $s_{AB}=+1$, $s_{BA}=-1$. Under order $o\in\{AB,BA\}$, define
\begin{equation}
  \xi_{tj}^{(o)}=(\theta_t+s_o\beta_j,\tau_t,-\theta_t-s_o\beta_j),
  \qquad
  p_{tj}^{L,(o)}=\softmax(\xi_{tj}^{(o)}),
  \label{eq:llm_logits}
\end{equation}
and the paired swap likelihood
\begin{equation}
  L_{tj}^{\rm swap}(\eta_t,\beta_j)
  = p_{tj}^{L,(AB)}(y_{tj}^{L,(AB)})\,p_{tj}^{L,(BA)}(y_{tj}^{L,(BA)}).
  \label{eq:compact_swaplike}
\end{equation}
The sign flip in $\xi_{tj}^{(o)}$ is the key structure: $\theta_t$ is order-invariant, whereas the judge bias enters with opposite signs.

\paragraph{Online target and streaming approximation.}
Let $\Phi=(\{C_r\}_{r\in\mathcal R},\{\beta_j\}_{j\in\mathcal J})$ denote the global evaluator parameters. The item likelihood factorizes as
\begin{equation}
  p(D_t\mid \eta_t,\Phi)
  = H_t(\eta_t,C)\prod_{j\in J_t}L_{tj}^{\rm swap}(\eta_t,\beta_j),
  \label{eq:item_likelihood}
\end{equation}
and, because $\eta_t$ is local while $\Phi$ is shared across the stream, the exact global posterior satisfies
\begin{equation}
  p(\Phi\mid D_{1:t}) \propto m_t(\Phi)\,p(\Phi\mid D_{1:t-1}),
  \qquad
  m_t(\Phi)=\int p(\eta_t)p(D_t\mid \eta_t,\Phi)\,d\eta_t.
  \label{eq:recursion}
\end{equation}
BayesJudge approximates this recursion with a Rao--Blackwellized assumed-density SMC filter. At time $t$, particles represent the joint judge-bias vector $\{\boldsymbol\beta_t^{(m)},w_t^{(m)}\}_{m=1}^M$, $\boldsymbol\beta_t^{(m)}=(\beta_{j,t}^{(m)})_{j\in\mathcal J}$, $\sum_{m=1}^M w_t^{(m)}=1$, and each human rater keeps Dirichlet sufficient statistics $\alpha_{r,k,\ell}^{(t)}$ with posterior mean $\widehat C_r^{(t)}[k,\ell]=\alpha_{r,k,\ell}^{(t)}/\sum_{\ell'}\alpha_{r,k,\ell'}^{(t)}$. For particle $m$, $\mathcal L_t^{(m)}(\eta)$ denotes the local log-posterior objective for item $t$, up to an additive constant:
\begin{equation}
  \mathcal L_t^{(m)}(\eta)
  = \log p(\eta)
  + \log H_t(\eta,\widehat C^{(t-1)})
  + \sum_{j\in J_t}\log L_{tj}^{\rm swap}(\eta,\beta_{j,t}^{(m)}),
  \label{eq:local_map_obj}
\end{equation}
where the first term is the item prior $p(\eta)=\mathcal N(\eta;\mu_0,\Sigma_0)$.
The implementation uses the MAP estimate $\widehat\eta_t^{(m)}=\arg\max_\eta \mathcal L_t^{(m)}(\eta)$ and a MAP-profile potential $\widehat m_t^{(m)}\propto \exp\{\mathcal L_t^{(m)}(\widehat\eta_t^{(m)})\}$ to update particle weights. The particle-weighted verdict distribution and the resulting hard Dirichlet update are
\begin{equation}
  \bar\pi_{t,k}=\sum_{m=1}^M w_t^{(m)} g_k(\hat\eta_t^{(m)}),
  \qquad
  k_t^\star=\arg\max_{k\in\mathcal K}\bar\pi_{t,k},
  \qquad
  \alpha_{r,k_t^\star,\ell}^{(t)}=\alpha_{r,k_t^\star,\ell}^{(t-1)}+\mathbb I[y_{tr}^H=\ell].
  \label{eq:responsibility}
\end{equation}

A soft variant distributing fractional counts $\bar\pi_{t,k}$ over rows is ablated in Appendix~\ref{app:compute}. Resampling is triggered when the effective sample size $\ESS_t<\kappa M$ for a threshold factor $\kappa\in(0,1)$, followed by a short joint move step for $\boldsymbol\beta$. At each step, BayesJudge outputs the plug-in verdict $\hat\pi_t=g(\hat\eta_t)$ over $\{A,T,B\}$ together with the running parameter estimates, whose entropy and values serve as uncertainty scores and evaluator diagnostics. Computational cost and pseudocode are in Appendices~\ref{app:compute} and~\ref{app:alg}.

\section{Protocol-Level Identifiability}
\label{sec:theory}

BayesJudge can only infer quantities that the data-collection protocol exposes. This section characterizes what each of the two protocol elements identifies: tie-open human labels expose tie mass, and paired AB/BA LLM calls expose position bias.

\begin{proposition}[Forced-choice labels do not identify tie mass]
\label{prop:tie_nonident}
Suppose the protocol censors the tie label and records only a forced choice between $A$ and $B$. We suppress the item index $t$ below. Let $q_A=\Pr(A\mid\{A,B\})\in(0,1)$. For every $u\in[0,1)$, the ternary distribution
\begin{equation}
\label{eq:forcedchoicefamily}
\pi_T=u,\qquad
\pi_A=(1-u)q_A,\qquad
\pi_B=(1-u)(1-q_A)
\end{equation}
induces the same forced-choice observation distribution. Hence $\pi_T$, and therefore the tie parameter $\tau$, is not identifiable from forced-choice-only data.
\end{proposition}

Proposition~\ref{prop:tie_nonident} assumes complete censoring of ties; our synthetic forced-choice ablation tests a milder corruption, randomly relabeling each true tie as $A$ or $B$ to mimic annotators forced to pick a side. Ambiguity remains unidentifiable: if $a$ is the probability that a true tie is recorded for the first response, the observed rate is $q=\pi_A+a\pi_T$, so all pairs $(\pi_A,\pi_T)$ satisfying this equation are observationally equivalent, and the data cannot separate genuine preference from absorbed tie mass (derivation in Appendix~\ref{app:random_folding}). Tie-open labels are therefore necessary whenever item-level ambiguity is part of the evaluation target.

\begin{proposition}[AB/BA swaps identify scalar position bias]
\label{prop:swap_ident}
Assume labels are expressed in canonical response coordinates, the only judge-side bias is scalar $\beta_j$, and $\ell_{AB},\ell_{BA}$ denote the log-odds of the judge's verdict probabilities. For the same item judged in both orders,
\begin{equation}
\ell_{AB}
=
\log\frac{p_{tj}^{L,(AB)}(A)}{p_{tj}^{L,(AB)}(B)}
=
2\theta_t+2\beta_j,
\qquad
\ell_{BA}
=
\log\frac{p_{tj}^{L,(BA)}(A)}{p_{tj}^{L,(BA)}(B)}
=
2\theta_t-2\beta_j.
\end{equation}
Therefore
\begin{equation}
\theta_t=\frac{\ell_{AB}+\ell_{BA}}{4},
\qquad
\beta_j=\frac{\ell_{AB}-\ell_{BA}}{4}.
\end{equation}
Note that $\tau_t$ cancels in the $A$-versus-$B$ log-odds, so this identification is unaffected by item ambiguity. With only one order, $\ell_{AB}=2(\theta_t+\beta_j)$, so item preference and presentation bias are confounded.
\end{proposition}

In experiments, LLMs provide categorical labels rather than explicit log-probabilities, so Proposition~\ref{prop:swap_ident} is a probability-level argument whose evidence accumulates across many swapped items. Both propositions are properties of the observation law and hold under any inference procedure; the concentration and calibration statements in Appendix~\ref{app:guarantees} assume exact inference, and the accuracy--cost trade-off of the implemented surrogate is evaluated in Appendix~\ref{app:inference_comparison}. Formal proofs appear in Appendix~\ref{app:proof}.
\section{Experiments}
\label{sec:experiments}

\subsection{Experimental Setup}
We validate BayesJudge on both synthetic and real-world benchmarks, with complementary goals. Synthetic data provide ground-truth evaluator parameters, allowing us to test parameter identifiability and structural correctness of the inference mechanism. Real-world data lack such ground truth, so we assess diagnostic validity and predictive reliability against observed human behavior. All datasets are split into 80/20 train/test partitions. Hyperparameters, synthetic dataset construction details, and SummEval conversion procedures are deferred to Appendices~\ref{app:dataset} and~\ref{app:sensitive}.

\begin{itemize}[leftmargin=*]
    \item \textbf{Synthetic Dataset.} A controlled benchmark of 1,000 pairwise comparisons annotated by 2 human raters (expert / novice) and 2 LLM judges (unbiased / position-biased) under AB/BA dual presentation orders. All human confusion matrices and LLM bias coefficients are prespecified by design, enabling ground-truth verification of parameter recovery and structural correctness.
     \item \textbf{Real-World Dataset (SummEval).} The original discrete-scoring dataset \citep{fabbri2021summeval} is converted into pairwise comparison format, yielding judgments from 8 human raters (3 experts, 5 crowdworkers) and GPT-4o-mini under dual presentation orders; the conversion procedure is detailed in Appendix~\ref{app:summeval}. We additionally evaluate BayesJudge on a broader model suite including Claude-Haiku-4.5 and DeepSeek-V4-Flash, with results in Appendix~\ref{app:multillm}.
    
\end{itemize}
\begin{wrapfigure}{r}{0.6\textwidth}
\centering
\includegraphics[width=1\linewidth]{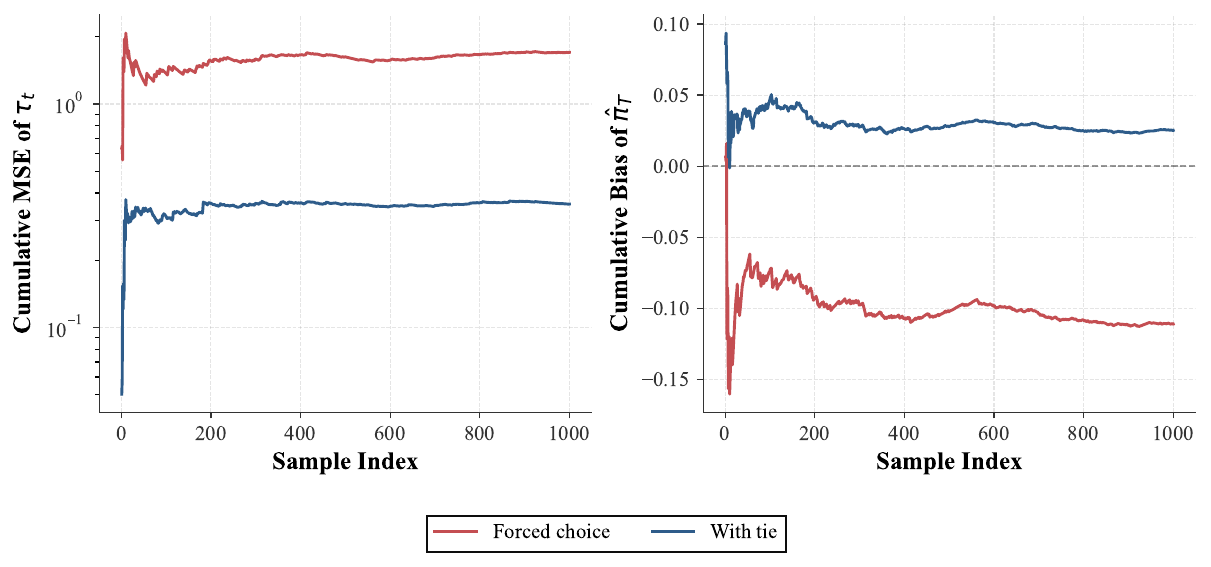}
\caption{Empirical validation of Proposition 1 on synthetic data. \textbf{Left:} Cumulative MSE of $\tau_t$ under forced-choice (red) vs. tie-open (blue) protocols. \textbf{Right:} Cumulative bias of posterior tie mass $\widehat{\pi}_{t,T}$. Without tie-open labels, tie mass remains an order of magnitude more erroneous and systematically biased; with them, the posterior converges to the generator's true ambiguity.}
\label{fig:tau_mse}
\end{wrapfigure}
\subsection{Results in Synthetic Dataset}
\paragraph{Observation 1: Tie-open labels are identification instruments for item-level ambiguity; without them, tie mass collapses to the prior.}

\begin{figure}[t]
\centering
\begin{minipage}[t]{0.49\linewidth}
  \centering
  \includegraphics[width=\linewidth]{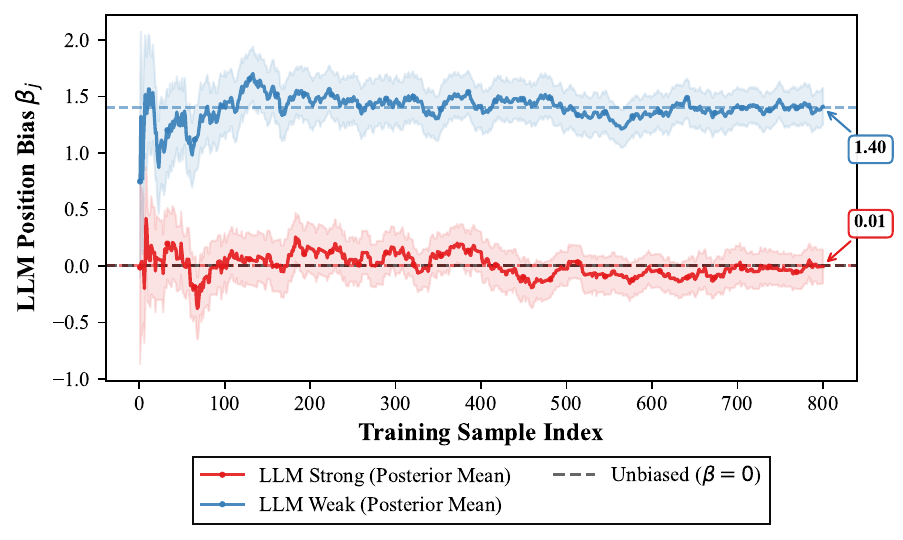}
\end{minipage}\hfill
\begin{minipage}[t]{0.49\linewidth}
  \centering
  \includegraphics[width=\linewidth]{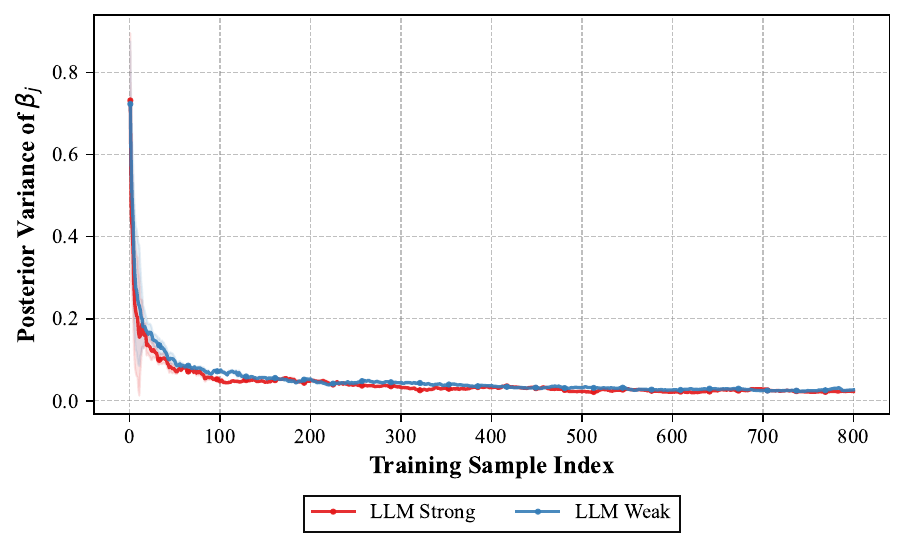}
\end{minipage}
\caption{Synthetic judge-bias recovery. Left: posterior means for an unbiased and a position-biased judge. Right: posterior-variance contraction over the same stream.}
\label{fig:beta_recovery}
\end{figure}

We validate Proposition~\ref{prop:tie_nonident} on a controlled synthetic stream by comparing two protocols under the same generator: one with open $\{A,T,B\}$ labels, and one where every $T$ label is randomly folded into a forced $A/B$ choice. Figure~\ref{fig:tau_mse} reports the cumulative MSE of $\tau_t$ (left) and the bias of $\widehat{\pi}_{t,T}$ (right). With tie labels, the MSE decays to roughly $10^{-1}$ and the bias vanishes. Under forced choice, the MSE stays an order of magnitude higher and the bias plateaus at $-0.1$ even after 1{,}000 samples. Tie mass is therefore not merely hard to estimate from forced-choice data; it is structurally unidentifiable, and the posterior tie probability falls back to the prior rather than recovering the true ambiguity. When item-level ambiguity matters, tie-open labels are a necessity, not an option.
\begin{wrapfigure}{r}{0.55\textwidth}
\centering
\includegraphics[width=0.78\linewidth]{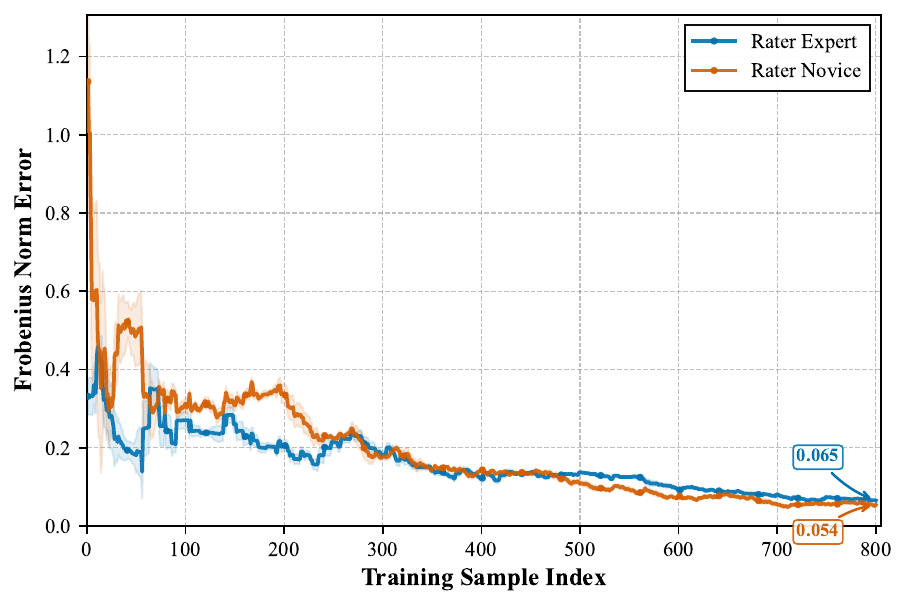}
\caption{Human confusion-matrix recovery on the completed synthetic stream. The curve reports Frobenius error between the current posterior mean of each rater matrix and the ground truth matrix.}
\label{fig:bias_recovery}
\end{wrapfigure}
\paragraph{Observation 2: BayesJudge accurately recovers ground-truth human confusion matrices in an online streaming setting and clearly identifies position bias in LLM evaluators.}
Figure~\ref{fig:bias_recovery} tracks the Frobenius-norm error between the recovered and ground-truth human confusion matrices after each sample. Final errors reach $0.065$ for the expert and $0.054$ for the novice, down from $0.332$ and $1.136$ at initialization. Figure~\ref{fig:beta_recovery} shows the posterior mean and variance of the LLM position-bias coefficient $\beta$ as samples accumulate: the variance shrinks monotonically, and the posterior mean converges to $0.01$ for the strong LLM (unbiased by design) versus $1.40$ for the weak LLM, whose positive value indicates a front-position bias, consistent with the ground-truth design of the two judges. Prior-initialization analyses for the confusion matrices appear in Appendix~\ref{app:sensitive}, and prior-sensitivity and initialization checks for $\beta$ in Appendix~\ref{app:human_confusion_analysis}; identifiability of $\beta$ on real-world judges such as GPT-4o-mini is analyzed in Appendix~\ref{app:supp}. Beyond point recovery, BayesJudge isolates an adversarial rater who always outputs the same label within that rater's $C_r$, leaving item ambiguity estimates unaffected (Appendix~\ref{app:adversarial}).

\begin{figure*}[t]
    \centering
    \begin{minipage}[b]{0.5\textwidth}
        \centering
        \includegraphics[height=4cm, width=1.15\linewidth]{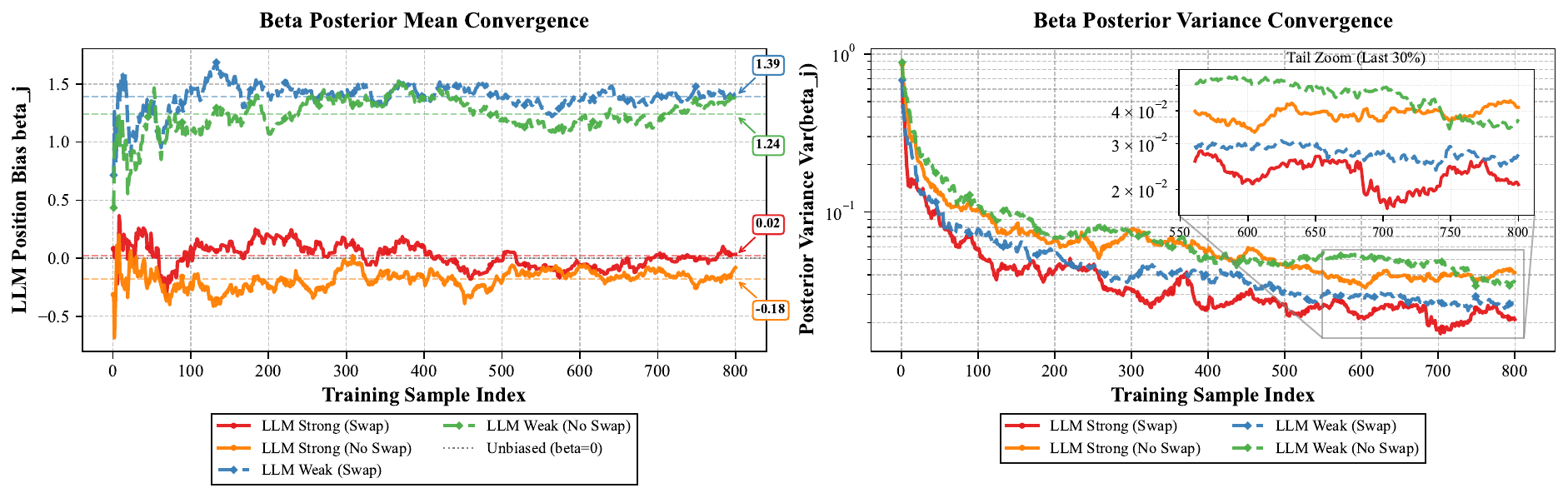}
        {\footnotesize (a) Bias mean and variance under swap vs. no-swap.}
    \end{minipage}
    \hfill
    \begin{minipage}[b]{0.43\textwidth}
        \centering
        \includegraphics[height=4cm, width=1\linewidth]{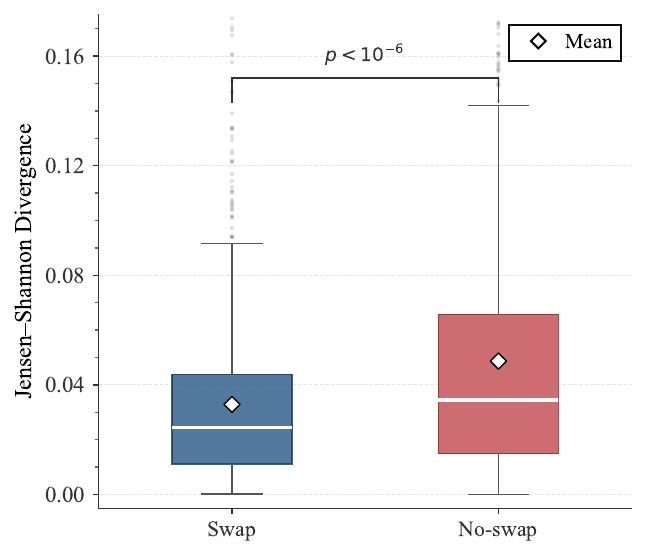}
        {\footnotesize (b) Per-item JS divergence; paired $t$-test $p<10^{-6}$.}
    \end{minipage}
    \caption{Ablation study on swap-based inference. Paired swaps produce less biased
    judge-bias estimates and tighter posterior agreement with the swap-identified posterior.}
    \label{fig:swap_compare}
\end{figure*}

\paragraph{Observation 3: Paired swaps are necessary to identify LLM position bias; without them, item preference and bias remain confounded and posterior uncertainty stays inflated.}
Figure~\ref{fig:swap_compare} compares the swap likelihood with the single-order likelihood within BayesJudge. With a single order, the posterior mean of $\beta$ remains markedly biased; with paired swaps, the estimate concentrates at the identifiable value, and its posterior variance converges much faster. This is the empirical counterpart of Proposition~\ref{prop:swap_ident}: because $\theta_t$ and $\beta_j$ enter the two orders with opposite signs (Eq.~\eqref{eq:llm_logits}), each swapped pair contributes complementary evidence that separates preference from bias, whereas single-order observations leave the two additive and unresolvable. SMC particle health diagnostics (ESS trajectories) are visualized in Appendix~\ref{app:ess}.

\paragraph{Observation 4: Jointly modeling human rater noise and LLM position bias is strictly better than freezing either channel.}
Table~\ref{tab:ablation} reports held-out predictive log-likelihood (PLL) on the synthetic test stream. The full model attains $-5.3356$ nats, versus $-6.0188$ when both channels are frozen; the gap of $0.6832$ nats corresponds to roughly a factor of two in average predictive probability. The gain is asymmetric across channels: freezing $\beta_j$ alone costs $0.5363$ nats, making position bias the dominant synthetic contaminant, whereas freezing the human update costs only $0.0264$ nats. Since the full model improves on both one-sided ablations, the two channels are complementary rather than interchangeable, although their contributions differ markedly in magnitude. PLL and ablation results on SummEval are reported in Appendix~\ref{app:supp}.
\begin{table}[t]
\centering
\caption{Ablation on held-out predictive log likelihood (PLL).}
\label{tab:ablation}
\begin{tabular}{lc}
\toprule
Model variant & PLL $\uparrow$ \\
\midrule
\textbf{BayesJudge} & $\mathbf{-5.3356}$ \\
w/o human rater update ($C_r=I$) & $-5.3620$ \\
w/o LLM bias update ($\beta_j=0$) & $-5.8719$ \\
w/o both updates & $-6.0188$ \\
\midrule
$\Delta$ (full vs. w/o both) & $+0.6832$ \\
\bottomrule
\end{tabular}
\end{table}

\subsection{Results in SummEval dataset}
\paragraph{Observation 5: Fusion induces an expert--crowd subgroup trade-off.}
We validate heterogeneous evaluator modeling via leave-one-human-out prediction on SummEval, with results in Table~\ref{tab:heldout} and the protocol detailed in Appendix~\ref{app:sensitive}. For each rater $r$, we withhold that rater's labels on the test split and predict them from the remaining human labels together with the LLM swap judgments, using the confusion matrix $\hat C_r$ calibrated on $r$'s training-split labels.

Comparing the full model against its single-channel variants reveals a subgroup trade-off. On experts, the full model attains the best accuracy at $65.3\%$, improving over both the LLM-only variant at $59.5\%$ and the human-only variant at $43.7\%$. On crowdworkers, it still beats the LLM-only variant in both accuracy and negative log-likelihood (NLL), but trails the human-only variant, whose consensus is dominated by the crowd majority. The baselines mirror this split. MACE~\citep{hovy2013learning} ties the human-only channel on crowdworkers at $69.1\%$ accuracy yet drops to $38.0\%$ on experts, where LLM evidence matters most. Bradley--Terry--Davidson~\citep{davidson1970ties} and GLAD~\citep{whitehill2009whose} sit between the two extremes on both subgroups without leading either. Overall, BayesJudge is the only method that leads on experts, MACE leads on crowdworkers, and no method wins both. The trade-off is structural. Joint training pulls the estimated rater confusion matrices toward the expert-aligned notion of quality carried by the LLM signal, whereas the human-only regime lets the crowd majority dominate the inferred consensus. We analyze the effect of panel imbalance in Appendix~\ref{app:supp}.

\begin{table}[t]
\centering
\footnotesize
\setlength{\tabcolsep}{3.5pt}
\caption{Held-out rater prediction on SummEval. The human channel predicts crowdworkers well, while the LLM channel captures expert judgment.}
\label{tab:heldout}
\begin{tabular}{@{}lcccc|cccc@{}}
\toprule
& \multicolumn{4}{c|}{\textbf{Experts (3 experts)}} & \multicolumn{4}{c}{\textbf{Crowdworkers (5 crowdworkers)}} \\
\cmidrule(lr){2-5} \cmidrule(l){6-9}
\textbf{Method} & \textbf{Acc$\uparrow$} & \textbf{NLL$\downarrow$} & \textbf{Brier$\downarrow$} & \textbf{ECE$\downarrow$} & \textbf{Acc$\uparrow$} & \textbf{NLL$\downarrow$} & \textbf{Brier$\downarrow$} & \textbf{ECE$\downarrow$} \\
\midrule
Majority-Vote & 44.8\% & 3.421 & 0.688 & 0.239 & 64.7\% & 2.034 & 0.511 & 0.078 \\
Bradley--Terry--Davidson & 46.33\% & 1.180 & 0.649 & 0.168 & 62.10\% & 1.111 & 0.531 & 0.066 \\
MACE & 38.00\% & 1.093 & 0.664 & 0.069 & \textbf{69.10}\% & \textbf{0.792} & \textbf{0.451} & \textbf{0.035} \\
GLAD & 44.33\% & 1.126 & 0.637 & 0.219 & 63.60\% & 1.773 & \underline{0.473} & 0.118 \\
\textbf{BayesJudge} & \textbf{65.3\%} & \textbf{0.886} & \textbf{0.515} & 0.109 & 42.3\% & 1.008 & 0.612 & \underline{0.042} \\
\textbf{BayesJudge(w/o Human)} & 59.5\% & 0.942 & 0.556 & 0.060 & 39.2\% & 1.075 & 0.653 & 0.053 \\
\textbf{BayesJudge(w/o LLM)} & 43.7\% & 1.019 & 0.621 & \textbf{0.056} & \underline{68.9\%} & \underline{0.854} & 0.491 & 0.176 \\
\bottomrule
\end{tabular}
\end{table}

\begin{figure}[t]
    \centering
    \begin{minipage}[t]{0.4\linewidth}
        \centering
        \vspace{0pt}
        \includegraphics[width=\linewidth]{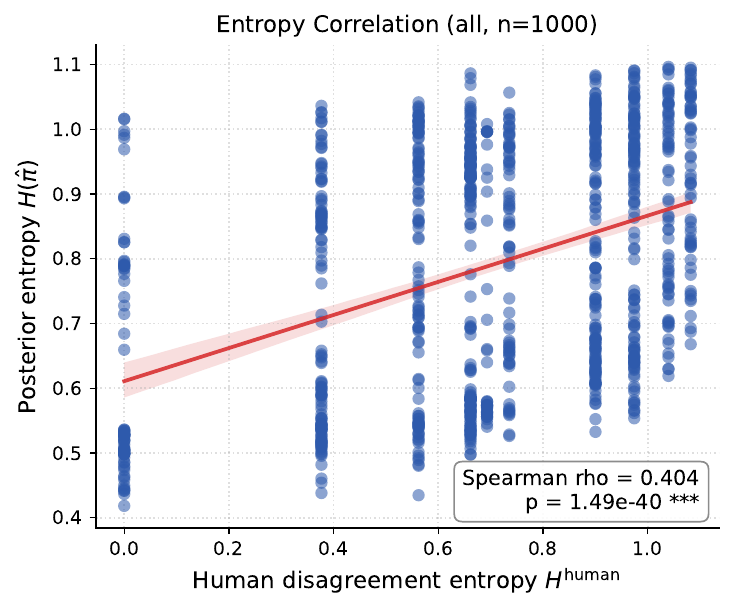}
        \caption{Entropy--disagreement correlation on SummEval. Higher model uncertainty tends to coincide with higher human disagreement.}
        \label{fig:entropy_correlation_all}
    \end{minipage}
    \hfill
    \begin{minipage}[t]{0.48\linewidth}
        \centering
        \vspace{0pt}
        \scriptsize
        \renewcommand{\arraystretch}{1.18}
        \setlength{\tabcolsep}{1.6pt}
        \captionof{table}{Quantile bucket analysis between posterior entropy $H(\widehat{\pi}_t)$ and human voting entropy $H^{\mathrm{human}}$ on SummEval ($n=1000$).}
        \label{tab:quantile}
        \vspace{0.4em}
        \begin{tabular}{@{}lcccc@{}}
            \toprule
            \textbf{Bucket} & \textbf{$H(\widehat{\pi}_t)$ range} & \textbf{$n$}
            & \textbf{Mean $H^{\mathrm{human}}$} & \textbf{Std $H^{\mathrm{human}}$} \\
            \midrule
            Q1 (lowest)  & $[0.418,0.574)$ & 200 & 0.4165 & 0.2848 \\
            Q2           & $[0.574,0.702)$ & 200 & 0.7757 & 0.2111 \\
            Q3           & $[0.702,0.895)$ & 200 & 0.7020 & 0.3275 \\
            Q4           & $[0.895,0.990)$ & 200 & 0.7441 & 0.2128 \\
            Q5 (highest) & $[0.990,1.096)$ & 200 & 0.8377 & 0.2291 \\
            \bottomrule
        \label{tab:entropy_buck}
        \end{tabular}
        \renewcommand{\arraystretch}{1.0}
    \end{minipage}
\end{figure}

\paragraph{Observation 6: Posterior entropy signals genuine human disagreement.}
Figure~\ref{fig:entropy_correlation_all} and Table~\ref{tab:entropy_buck} compare posterior entropy against human voting entropy across the eight annotators. The two correlate positively, with Spearman $\rho=0.404$ at $p\approx1.5\times10^{-40}$. The separation across quantiles is sharp: mean human disagreement rises from $0.4165$ in the lowest posterior-entropy quintile to $0.8377$ in the highest, that is, from about $38\%$ to $76\%$ of the maximal ternary entropy $\ln 3 \approx 1.10$. The steepest rise occurs between the first two quintiles, so even a modest increase in posterior entropy already marks substantially more ambiguous items. The trend is not perfectly monotone in the middle buckets, and for an instructive reason: BayesJudge conditions on LLM swap evidence and inferred rater reliability, so it can assign high entropy to items where humans look unanimous but swapped judgments expose hidden ambiguity, or discount apparent disagreement that traces back to unreliable raters rather than genuinely indistinguishable items.

\subsection{Summary}
The above experiments validate three claims. First, on synthetic data, BayesJudge recovers ground-truth evaluator parameters, confirming that tie-open labels and AB/BA swaps serve as protocol-level identification instruments. Second, on SummEval, the inferred confusion matrices separate expert and crowdworker signatures without rater metadata, and fusion yields a substantial gain on expert prediction while the human-only variant retains the advantage on the crowd subgroup, a structural trade-off rather than uniformly beneficial fusion. Third, posterior entropy correlates with real human disagreement, yielding actionable uncertainty signals for routing. Together, these results show that conflicting and biased judgments, once modeled rather than collapsed, become interpretable diagnostics of the evaluation process itself. We additionally evaluate the same protocol on MT-Bench, with results in Appendix~\ref{app:mtbench}.

\section{Limitations}
\label{sec:discussion}
On existing real-world evaluation datasets, each item receives only a handful of human annotations, and the annotator set varies across items, which limits how precisely BayesJudge can estimate individual evaluator bias patterns. Second, we model only LLM position bias for analytical tractability, while practical judgments are shaped by many other factors such as response length. An antisymmetric covariate extension that accommodates such factors is sketched in Appendix~\ref{app: extension}. Third, we assume a fixed evaluator pool with stationary behavior, hence the framework is best suited to relatively homogeneous tasks with repeated observations per rater. Finally, the downstream utility of the posterior verdict distributions, for example in routing or training-data curation, remains to be validated directly, and we leave this to future work.

\section{Conclusion}
We propose BayesJudge, an online Bayesian meta-evaluation framework that jointly infers item ambiguity, human rater confusion, and LLM position bias from conflicting judgment streams. Synthetic experiments show that BayesJudge recovers evaluator parameters online, that tie-open labels and paired swaps are necessary for identifiability, and that jointly modeling both bias channels outperforms freezing either one. On the SummEval dataset, the framework separates expert and crowdworker signatures without rater metadata, detects systematic position effects in LLM judgments, and yields uncertainty estimates that correlate with human disagreement. We hope that moving evaluation verdicts from hard labels to posterior distributions encourages the community to treat evaluator disagreement as a signal worth modeling in evaluation practice.

\begin{ack}
  We thank the anonymous reviewers for their constructive feedback. The work is supported by grants from Guangdong Special Funding for Science and Technology Innovation Strategy ( 2024A1111120004).

\end{ack}

\bibliographystyle{plainnat}
\bibliography{bayesjudge_refs}
\newpage
\appendix
\section{Appendix}
The appendix of this paper is organized as follows:
\begin{itemize}
    \item \textbf{Appendix \ref{app:dataset}.} Dataset Details.
    \item \textbf{Appendix \ref{app:proof}.} Theoretical Proofs and Additional Derivations.
    \item \textbf{Appendix \ref{app:sensitive}.} Hyperparameter Setting And Sensitivity Analysis.
    \item \textbf{Appendix \ref{app:ess}.} SMC Particle Health Diagnostics.
    \item \textbf{Appendix \ref{app:compute}.} Computational Efficiency.
    
    \item \textbf{Appendix \ref{app:alg}.} Algorithm Pseudocode.
    \item \textbf{Appendix \ref{app:prompt}.} Prompt.
    \item \textbf{Appendix \ref{app:supp}.} Supplementary Analysis.

    \item \textbf{Appendix~\ref{app:mtbench}.} MT-Bench Evaluation.
\end{itemize}

\section{Dataset Details}
\label{app:dataset}
\subsection{Synthetic Data Generation Details}
\label{app:synthetic}
We construct a controlled synthetic dataset of 1,000 pairwise comparison samples with an 80/20 chronological split (800 training / 200 test). Each sample is annotated by 2 human raters and 2 LLM judges. Ground-truth latent parameters are sampled to yield balanced outcome categories (A wins / Tie / B wins at 40\% / 20\% / 40\%).

\subsubsection{Ground-Truth Parameterization}

Each sample $t$ is endowed with a latent quality vector $\eta_t = (\theta_t, \tau_t) \in \mathbb{R}^2$. The quality-difference parameter $\theta_t$ encodes the intrinsic quality advantage of model A over model B ($\theta_t > 0$ favors A; $\theta_t < 0$ favors B; $\theta_t \approx 0$ indicates near-parity). The tie-propensity parameter $\tau_t$ governs the ambiguity of the comparison; larger values increase the probability of a tie outcome.

To ensure balanced category frequencies, we sample $\eta_t$ from group-conditional Gaussian distributions:

\begin{itemize}
    \item \textbf{A-win group} (40\%): $\theta_t \sim \mathcal{N}(2.0, 0.5)$, $\tau_t \sim \mathcal{N}(0.1, 0.05)$;
    \item \textbf{B-win group} (40\%): $\theta_t \sim \mathcal{N}(-2.0, 0.5)$, $\tau_t \sim \mathcal{N}(0.1, 0.05)$;
    \item \textbf{Tie group} (20\%): $\theta_t \sim \mathcal{N}(0.0, 0.3)$, $\tau_t \sim \mathcal{N}(1.5, 0.3)$.
\end{itemize}

After generation, all samples are randomly shuffled to eliminate group ordering. The ground-truth categorical distribution $\pi_t \in \Delta^2$ is obtained via softmax:
\[
\pi_t = \text{softmax}\bigl([\theta_t,\; \tau_t,\; -\theta_t]\bigr),
\]
where the three dimensions correspond to \{A wins, Tie, B wins\}, indexed by $\{0, 1, 2\}$. The true label $y_t^*$ is determined by $y_t^* = \arg\max_{k} \pi_{t,k}$.

\subsubsection{Human Rater Simulation}
Each human rater $r$ is characterized by a row-stochastic confusion matrix $C_r \in [0,1]^{3 \times 3}$, where entry $C_r[i,j]$ denotes the conditional probability $p(y^{\text{obs}} = j \mid y^* = i)$. We instantiate two prototypical skill levels:

\paragraph{Expert rater (Rater 0).} Near-diagonal cognition with high accuracy:
\[
C_0 = 
\begin{bmatrix}
0.90 & 0.05 & 0.05 \\
0.10 & 0.80 & 0.10 \\
0.10 & 0.05 & 0.85
\end{bmatrix}.
\]

\paragraph{Novice rater (Rater 1).} Noisy, diffuse judgment with systematic bias toward over-selecting A:
\[
C_1 = 
\begin{bmatrix}
0.50 & 0.25 & 0.25 \\
0.30 & 0.40 & 0.30 \\
0.25 & 0.25 & 0.50
\end{bmatrix}.
\]

For each sample $t$, let $i = y_t^*$ denote the true label index. Each rater's observed label is sampled independently from the categorical distribution defined by the $i$-th row of $C_r$.

\subsubsection{LLM Judge Simulation under Swap Design}

Following the swap design formalism in Section~\ref{sec:model}, each LLM judge $j$ is parameterized by a scalar position-bias coefficient $\beta_j \in \mathbb{R}$. Positive $\beta_j$ induces a front-position bias (favoring the item presented first); $\beta_j = 0$ indicates position invariance.

Given sample $t$ with latent parameters $(\theta_t, \tau_t)$, the logits under the two presentation orders are:

\textbf{AB order} (A in position 1, B in position 2):
\[
\ell_t^{(AB)} = [\theta_t + \beta_j,\; \tau_t,\; -\theta_t - \beta_j]^\top,
\]

\textbf{BA order} (B in position 1, A in position 2):
\[
\ell_t^{(BA)} = [\theta_t - \beta_j,\; \tau_t,\; - \theta_t + \beta_j]^\top,
\]

where the three components correspond to \{A, Tie, B\}, respectively. The observed judgments are obtained by independent softmax sampling:
\[
y_t^{(AB)} \sim \text{Categorical}\bigl(\text{softmax}(\ell_t^{(AB)})\bigr), \quad
y_t^{(BA)} \sim \text{Categorical}\bigl(\text{softmax}(\ell_t^{(BA)})\bigr).
\]

We instantiate two LLM profiles:
\begin{itemize}
    \item \textbf{Strong judge:} $\beta = 0.0$ (no position bias, decisions depend solely on intrinsic quality);
    \item \textbf{Weak judge:} $\beta = 2.0$ (strong front-position bias, heavily influenced by presentation order).
\end{itemize}

The paired judgments $(y_t^{(AB)}, y_t^{(BA)})$ provide a principled signal for identifying and correcting position bias without requiring ground-truth labels.

\subsection{SummEval Conversion Details}
\label{app:summeval}

\paragraph{Raw data format.} The original SummEval corpus \citep{fabbri2021summeval} contains 100 news documents from CNN/DailyMail, each summarized by 16 systems. It comprises a collection of records $\mathcal{D}_{\text{raw}} = \{r_k\}_{k=1}^{N}$, where each record corresponds to a single machine-generated summary. Formally, a record is a tuple
\[
r_k = \langle q_k,\, x_k,\, m_k,\, y_k,\, \mathcal{A}_k^{\text{exp}},\, \mathcal{A}_k^{\text{turk}} \rangle,
\]
with the following components: $q_k \in \mathcal{Q}$ is a source-document identifier; $x_k \in \mathcal{X}$ denotes the source text; $m_k \in \mathcal{M}$ identifies the summarization model; $y_k \in \mathcal{Y}$ is the generated summary; $\mathcal{A}_k^{\text{exp}} = \{a_{k,i}^{\text{exp}}\}_{i=1}^{R_e}$ collects expert annotations ($R_e=3$); and $\mathcal{A}_k^{\text{turk}} = \{a_{k,i}^{\text{turk}}\}_{i=1}^{R_t}$ collects crowdworker annotations ($R_t=5$). Each individual annotation is a $d$-dimensional quality vector $a_{k,i} = (a_{k,i}^{(1)}, \dots, a_{k,i}^{(d)}) \in \mathbb{R}^d$.

\paragraph{Grouping by source document.} To ensure that comparisons are grounded on identical source material, we partition $\mathcal{D}_{\text{raw}}$ into groups of records sharing the same source text. Define the grouping operator $\mathcal{G}: \mathcal{D}_{\text{raw}} \to 2^{\mathcal{D}_{\text{raw}}}$ such that each group $\mathcal{S}_g$ satisfies
\[
\mathcal{S}_g = \{r \in \mathcal{D}_{\text{raw}} \mid x = x_g\}, \qquad |\mathcal{S}_g| = G,
\]
where $G=16$ is the fixed group size (number of distinct models per source). Let $T = |\mathcal{G}(\mathcal{D}_{\text{raw}})|$ denote the number of valid source-document groups.

\paragraph{Pairwise enumeration.} Within each group $\mathcal{S}_g = \{r_{g,1}, \dots, r_{g,G}\}$, we enumerate all unordered model pairs via the index set
\[
\mathcal{P}_g = \{(i,j) \in [G] \times [G] \mid i < j\},
\]
yielding $|\mathcal{P}_g| = \binom{G}{2} = 120$ distinct pairwise comparisons per source. For each pair $(i,j) \in \mathcal{P}_g$, we construct a comparison sample $s_g^{(i,j)}$ with candidate A drawn from model $m_{g,i}$ and candidate B from model $m_{g,j}$. The A/B assignment is deterministically fixed by the original model index ($i < j$) to prevent random order effects and ensure reproducibility across experimental runs.

\paragraph{Score aggregation.} Human annotators provide multi-dimensional quality assessments: each annotation rates a summary on four dimensions (coherence, consistency, fluency, relevance) on a 1--5 Likert scale. To obtain a scalar quality score for each candidate under each rater, we apply a dimension-wise averaging operator $\phi: \mathbb{R}^d \to \mathbb{R}$ with $d = 4$:
\[
\phi(a) = \frac{1}{d}\sum_{\ell=1}^{d} a^{(\ell)}.
\]
For a given pair $(i,j)$ and annotator $r$, the scalar scores for candidates A and B are
\[
u_{g,r}^{(A)} = \phi\bigl(a_{g,r}^{(i)}\bigr), \qquad u_{g,r}^{(B)} = \phi\bigl(a_{g,r}^{(j)}\bigr).
\]

\paragraph{Label discretization.} We convert continuous scalar scores into categorical pairwise judgments via a thresholded comparator $\psi: \mathbb{R} \times \mathbb{R} \to \{\mathsf{A}, \mathsf{B}, \mathsf{T}\}$:
\[
\psi\bigl(u^{(A)}, u^{(B)}\bigr) = 
\begin{cases}
\mathsf{T}, & \text{if } |u^{(A)} - u^{(B)}| < \varepsilon, \\[4pt]
\mathsf{A}, & \text{if } u^{(A)} > u^{(B)} + \varepsilon, \\[4pt]
\mathsf{B}, & \text{if } u^{(B)} > u^{(A)} + \varepsilon,
\end{cases}
\]
where $\varepsilon = 10^{-6}$ is the tie tolerance. The resulting label alphabet $\{\mathsf{A}, \mathsf{B}, \mathsf{T}\}$ denotes ``A wins,'' ``B wins,'' and ``Tie,'' respectively.

\paragraph{Annotator-consistency constraint.} A critical design choice is that each pairwise label is derived from the \emph{same} annotator's evaluation of both candidates. Formally, for group $g$, pair $(i,j)$, and annotator index $r \in \mathcal{R}_e \cup \mathcal{R}_t$ (where $\mathcal{R}_e = \{1,\dots,R_e\}$ and $\mathcal{R}_t = \{1,\dots,R_t\}$), the observed label is
\[
w_{g,r}^{(i,j)} = 
\begin{cases}
\psi\bigl(\phi(a_{g,r}^{\text{exp},(i)}),\, \phi(a_{g,r}^{\text{exp},(j)})\bigr), & r \in \mathcal{R}_e, \\[6pt]
\psi\bigl(\phi(a_{g,r}^{\text{turk},(i)}),\, \phi(a_{g,r}^{\text{turk},(j)})\bigr), & r \in \mathcal{R}_t.
\end{cases}
\]
This within-rater pairing eliminates inter-rater calibration bias from the comparison signal: any systematic leniency or strictness of annotator $r$ affects both candidates equally and thus cancels out in the relative judgment.

\paragraph{Output structure.} The final pairwise dataset is $\mathcal{D}_{\text{pair}} = \{s_m\}_{m=1}^{M}$, where $M = T \cdot \binom{G}{2}$. Each sample $s_m$ is a structured tuple
\[
s_m = \langle u_m,\, q_m,\, m_m^{(A)},\, m_m^{(B)},\, x_m,\, y_m^{(A)},\, y_m^{(B)},\, \mathcal{J}_m,\, \mathbf{w}_m,\, \mathcal{A}_m^{\text{det}} \rangle,
\]
with components defined as follows: $u_m$ is a global unique identifier; $q_m$ the source-document ID; $m_m^{(A)}, m_m^{(B)}$ the competing model IDs; $x_m$ the shared source text; $y_m^{(A)}, y_m^{(B)}$ the candidate summaries; $\mathcal{J}_m$ the sequence of annotator IDs; $\mathbf{w}_m \in \{\mathsf{A}, \mathsf{B}, \mathsf{T}\}^{R_e+R_t}$ the corresponding label vector; and $\mathcal{A}_m^{\text{det}}$ the raw $d$-dimensional scores retained for diagnostic traceability.

\paragraph{Dataset statistics.} With $G=16$ models per source document, each group yields $\binom{16}{2}=120$ pairwise samples. Aggregating over $T$ source documents produces $M = 120T$ total pairwise comparisons, every sample carrying $R_e + R_t = 8$ independent human judgments. For the main experiments, we uniformly subsample 1{,}000 pairs stratified across the $T = 100$ source documents, yielding 8{,}000 human judgments in total (Table~\ref{tab:label_dist}). The tie threshold $\varepsilon$ is chosen sufficiently small so that $\mathsf{T}$ labels arise only from near-exact score equality; consequently, ties are sparse in practice and reflect genuine quality indistinguishability rather than numerical rounding artifacts.

To validate this conversion empirically, we examine the SummEval dataset after transformation. Pitting each system against every other yields 8,000 human judgments across 1,000 sample pairs rated by 8 annotators. The resulting label distribution shows a balanced preference split (40.39\% vs. 44.92\%) with a moderate tie rate of 14.69\%, indicating that the conversion preserves meaningful discriminative signal rather than collapsing into unanimous or random outcomes. Moreover, tie usage remains consistent across rater demographics: experts select Tie at 15.93\% while Turkers do so at 13.94\%---a negligible gap of merely 2.0 percentage points. This near-identical frequency suggests both populations interpret the tie option similarly, confirming that the pairwise conversion does not introduce group-specific response artifacts and remains interpretable across diverse annotator backgrounds.

\begin{table}[h]
\centering
\caption{Overall label distribution of SummEval in pairwise-comparison format (8,000 human judgments = 1,000 samples $\times$ 8 raters).}
\label{tab:label_dist}
\begin{tabular}{@{}lcc@{}}
\toprule
\textbf{Label} & \textbf{Count} & \textbf{Percentage} \\
\midrule
A (Model A wins) & 3,231 & 40.39\% \\
B (Model B wins) & 3,594 & 44.92\% \\
Tie & 1,175 & 14.69\% \\
\midrule
\textbf{Total} & \textbf{8,000} & \textbf{100.00\%} \\
\bottomrule
\end{tabular}
\end{table}

\begin{table}[h]
\centering
\caption{Tie usage comparison between Expert and Turker raters.}
\label{tab:expert_turker}
\begin{tabular}{@{}lccc@{}}
\toprule
\textbf{Group} & \textbf{Total Judgments} & \textbf{Tie Count} & \textbf{Tie Rate} \\
\midrule
Expert (3 raters) & 3,000 & 478 & 15.93\% \\
Turker (5 raters) & 5,000 & 697 & 13.94\% \\
\bottomrule
\end{tabular}
\end{table}

\section{Theoretical Proofs and Additional Derivations}
\label{app:proof}

\subsection{Full joint posterior}
The augmented item-level model uses rater-specific latent verdicts
$z_{tr}\in\mathcal K$ for the human channel:
\begin{align}
  z_{tr}\mid \eta_t &\sim \Cat(g(\eta_t)), \\
  y_{tr}^H\mid z_{tr}=k,C_r &\sim \Cat(C_r[k,:]), \\
  y_{tj}^{L,(o)}\mid \eta_t,\beta_j
  &\sim \Cat(p_{tj}^{L,(o)}),\qquad o\in\{AB,BA\}.
\end{align}
After marginalizing $z_{tr}$, the item likelihood is Equation~\eqref{eq:item_likelihood}. For completeness, the corresponding posterior over local item variables and global evaluator parameters is
\begin{align}
  p(\eta_{1:t},\Phi\mid D_{1:t})
  &\propto
  \prod_{\tau=1}^t
  \Bigg[
    \mathcal N(\eta_\tau;\mu_0,\Sigma_0)
    H_\tau(\eta_\tau,C)
    \prod_{j\in J_\tau}L_{\tau j}^{\rm swap}(\eta_\tau,\beta_j)
  \Bigg] \notag\\
  &\quad\times
  \prod_{r\in\mathcal R}\prod_{k\in\mathcal K}
  \Dir\!\left(C_r[k,:];\lambda_{\rm off}\mathbf 1+(\lambda_{\rm diag}-\lambda_{\rm off})e_k\right)
  \prod_{j\in\mathcal J}\mathcal N(\beta_j;0,\sigma_\beta^2),
  \label{eq:full_joint_posterior}
\end{align}
where
\begin{equation}
H_\tau(\eta_\tau, C) = \prod_{r\in R_\tau}\sum_{k\in\mathcal K} g_k(\eta_\tau)\, C_r[k, y_{\tau r}^H].
\end{equation}
Marginalizing the new local variable $\eta_t$ gives the online recursion in Equation~\eqref{eq:recursion}.

\subsection{Proof of Proposition~\ref{prop:tie_nonident}}
For forced-choice-only observations, the protocol records $q_A=P(A\mid A\cup B)$. For any $u\in[0,1)$, define the ternary distribution in Equation~\eqref{eq:forcedchoicefamily}. Then
\begin{align}
  \frac{\pi_A}{\pi_A+\pi_B}
  =\frac{(1-u)q_A}{(1-u)q_A+(1-u)(1-q_A)}=q_A.
\end{align}
Thus the full one-parameter family indexed by $u$ induces the same forced-choice distribution. No estimator can recover $u$ without tie-open labels or other auxiliary information.

\subsection{Random-folding variant used in the synthetic ablation}
\label{app:random_folding}

The synthetic forced-choice ablation randomly folds true tie labels into
forced \(A/B\) observations. Suppose
\[
P(\widetilde A\mid T)=a,\qquad
P(\widetilde B\mid T)=1-a,
\qquad a\in(0,1).
\]
The observed forced-choice probability is
\[
q=P(\widetilde A)=P(A)+P(\widetilde A\mid T)P(T)
=\pi_A+a\pi_T.
\]
For any
\[
0\le u\le \min\{q/a,(1-q)/(1-a)\},
\]
define
\[
\pi_T=u,\qquad
\pi_A=q-au,\qquad
\pi_B=1-q-(1-a)u.
\]
Then
\[
\pi_A+a\pi_T=(q-au)+au=q,
\]
and
\[
\pi_A+\pi_T+\pi_B
=
(q-au)+u+1-q-(1-a)u
=
1.
\]
The stated range of \(u\) ensures non-negativity. Hence a continuum of
different tie masses induces the same observed forced-choice probability
\(q\). Randomly folded forced-choice labels therefore cannot identify
\(\pi_T\) without tie-open labels or other auxiliary information.

\subsection{Proof of Proposition~\ref{prop:swap_ident}}
The proof requires the canonical-coordinate convention in Section~\ref{sec:model}. Under the scalar swap model, the A/B probability log-odds in $AB$ order are
\begin{align}
  \ell_{AB}
  =\log\frac{\exp(\theta_t+\beta_j)}{\exp(-\theta_t-\beta_j)}
  =2\theta_t+2\beta_j,
\end{align}
and in $BA$ order they are
\begin{align}
  \ell_{BA}
  =\log\frac{\exp(\theta_t-\beta_j)}{\exp(-\theta_t+\beta_j)}
  =2\theta_t-2\beta_j.
\end{align}
Adding and subtracting these equations then completes the proof. With only $AB$ observations, the likelihood depends on $\theta_t+\beta_j$; therefore $\theta_t'=\theta_t+c$ and $\beta_j'=\beta_j-c$ yield the same A/B log-odds for any constant $c$, proving the confounding claim.

\subsection{Collapsed human likelihood and responsibilities}
Starting from the augmented human channel, where each rater's label is
generated from its own latent verdict $z_{tr}\mid\eta_t\sim\mathrm{Cat}(g(\eta_t))$,
\begin{align}
  p(Y_t^H\mid \eta_t,C)
  &=\prod_{r\in R_t}\sum_{k\in\mathcal K}p(z_{tr}=k\mid\eta_t)\,p(y_{tr}^H\mid z_{tr}=k,C_r) \\
  &=\prod_{r\in R_t}\sum_{k\in\mathcal K}g_k(\eta_t)\,C_r[k,y_{tr}^H],
\end{align}
which is Equation~\eqref{eq:human_joint_likelihood}. The conditional posterior
responsibility of each rater's latent verdict is therefore
\begin{align}
  p(z_{tr}=k\mid y_{tr}^H,\eta_t,C)
  =\frac{g_k(\eta_t)\,C_r[k,y_{tr}^H]}
  {\sum_{h\in\mathcal K}g_h(\eta_t)\,C_r[h,y_{tr}^H]}.
  \label{eq:responsibilities}
\end{align}
An exact conjugate update would assign (hard) or distribute (soft) the per-rater responsibility in Eq.~\ref{eq:responsibilities}. The streaming implementation instead uses the label-free particle-weighted verdict distribution $\bar\pi_t$ (Eq.~\ref{eq:responsibility}, which is shared across raters.

\subsection{Swap likelihood in expanded form}
For judge $j$ and item $t$,
\begin{align}
  L_{tj}^{\rm swap}(\eta_t,\beta_j)
  =\frac{\exp(\xi_{tj,y_{tj}^{L,(AB)}}^{(AB)})}{\sum_{k\in\mathcal K}\exp(\xi_{tj,k}^{(AB)})}
   \cdot
   \frac{\exp(\xi_{tj,y_{tj}^{L,(BA)}}^{(BA)})}{\sum_{k\in\mathcal K}\exp(\xi_{tj,k}^{(BA)})}.
\end{align}
The sign of $\beta_j$ flips across the two factors while $\theta_t$ does not, so the paired likelihood contains both symmetric and antisymmetric information.

\subsection{Local item update and particle weights}
Given joint particle $m$, the local MAP objective and the profile potential used by the streaming implementation are
\begin{align}
  \mathcal L_t^{(m)}(\eta)
  &=\log\mathcal N(\eta;\mu_0,\Sigma_0)
    +\log H_t(\eta,\widehat C^{(t-1)})
    +\sum_{j\in J_t}\log L_{tj}^{\rm swap}(\eta,\beta_{j,t-1}^{(m)}),
    \label{eq:localobj}\\[-0.5ex]
  \widehat m_t^{(m)}
  &=\exp\{\mathcal L_t^{(m)}(\widehat\eta_t^{(m)})\},
    \qquad \widehat\eta_t^{(m)}=\arg\max_\eta \mathcal L_t^{(m)}(\eta).
    \label{eq:profile_potential}
\end{align}
An optional two-dimensional Laplace correction is
\begin{align}
  \widehat m_{t,{\rm Lap}}^{(m)}=(2\pi)^{d/2}|\mathcal H_t^{(m)}|^{-1/2}
  \exp\{\mathcal L_t^{(m)}(\widehat\eta_t^{(m)})\},\quad
  \mathcal H_t^{(m)}=-\nabla_\eta^2\mathcal L_t^{(m)}(\widehat\eta_t^{(m)}),\ d=2.
  \label{eq:laplace_potential}
\end{align}
If a short random-walk Metropolis--Hastings chain is used instead of direct optimization, a symmetric proposal $q(\eta^\star\mid\eta)=q(\eta\mid\eta^\star)$ has acceptance ratio
\begin{align}
  a=\min\Bigg\{1,
  \frac{\mathcal N(\eta^\star;\mu_0,\Sigma_0)
  H_t(\eta^\star,\widehat C^{(t-1)})
  \prod_{j\in J_t}L_{tj}^{\rm swap}(\eta^\star,\beta_j^{(m)})}
  {\mathcal N(\eta;\mu_0,\Sigma_0)
  H_t(\eta,\widehat C^{(t-1)})
  \prod_{j\in J_t}L_{tj}^{\rm swap}(\eta,\beta_j^{(m)})}
  \Bigg\}.
\end{align}
The exact incremental potential is the marginal likelihood $m_t(\Phi)$ in Equation~\eqref{eq:recursion}. The streaming implementation uses the profile approximation in Equation~\eqref{eq:profile_potential}; Equation~\eqref{eq:laplace_potential} gives the optional Laplace correction. Consequently, the implemented particle system targets an assumed-density surrogate posterior rather than the exact posterior.

\subsection{Resampling and rejuvenation}
Particle degeneracy is monitored by
\begin{align}
  \ESS_t=\frac{1}{\sum_{m=1}^M (w_t^{(m)})^2}.
\end{align}
If $\ESS_t<\kappa M$, systematic resampling is applied to the joint particles $\boldsymbol\beta_t^{(m)}$, not separately to each judge coordinate. A resample-move step proposes
\begin{align}
  \boldsymbol\beta^\star\sim\mathcal N(\boldsymbol\beta_t^{(m)},\epsilon_t^2 I),
  \qquad
  \epsilon_t=\epsilon_0/\sqrt t.
\end{align}
For a buffer $\mathcal B_t$ of recent items, the corresponding approximate Metropolis ratio is
\begin{align}
  a=\min\Bigg\{1,
  \frac{\mathcal N(\boldsymbol\beta^\star;0,\sigma_\beta^2I)
  \prod_{\tau\in\mathcal B_t}\widehat m_\tau(\boldsymbol\beta^\star,\widehat C^{(\tau-1)})}
  {\mathcal N(\boldsymbol\beta;0,\sigma_\beta^2I)
  \prod_{\tau\in\mathcal B_t}\widehat m_\tau(\boldsymbol\beta,\widehat C^{(\tau-1)})}
  \Bigg\}.
\end{align}
A longer buffer better approximates the accumulated surrogate posterior; a short buffer is cheaper and was used in the streaming experiments.

\subsection{Human confusion updates}
The exact augmented model would sample $z_{tr}$ independently for each
observed rater $r$ and update row $z_{tr}$ of that rater's confusion
matrix by a conjugate Dirichlet increment. The streaming implementation instead uses assignment weights
\begin{align}
  q_{t,k}=\mathbb I[k=\arg\max_h\bar\pi_{t,h}]\quad\text{(hard, default)},
  \qquad q_{t,k}=\bar\pi_{t,k}\quad\text{(soft ablation)},
  \label{eq:hard_assignment}
\end{align}
followed by the Dirichlet pseudo-count update
\begin{align}
  \alpha_{r,k,\ell}^{(t)}=\alpha_{r,k,\ell}^{(t-1)}+
  \mathbb I[y_{tr}^H=\ell]q_{t,k},\qquad r\in R_t.
  \label{eq:confusion_update}
\end{align}
Both updates are assumed-density approximations, not full Gibbs updates.

\subsection{Additional theoretical guarantees}
\label{app:guarantees}
\begin{proposition}[Posterior concentration of global judge bias]
Assume the data are generated by the exact model, the swap likelihood is locally identifiable, the prior density for $\beta_j$ is positive in a neighborhood of the true value $\beta_j^\star$, and the incremental likelihood satisfies standard local asymptotic normality conditions. Then the exact posterior for $\beta_j$ is consistent and
\begin{align}
  \Var(\beta_j\mid D_{1:t})=O(t^{-1}).
\end{align}
Equivalently, the posterior uncertainty radius contracts at rate $O_p(t^{-1/2})$. More precisely, the centered posterior is asymptotically normal with variance given by the inverse accumulated Fisher information.
\end{proposition}

\begin{proposition}[SMC approximation error for fixed horizon]
Fix $t$ and let $f$ be any bounded measurable function of $\boldsymbol\beta$. For ideal SMC with exact incremental potentials and regular resampling/move kernels invariant to the exact target, $\sum_{m=1}^M w_t^{(m)}f(\boldsymbol\beta_t^{(m)})$ converges in probability to $\mathbb E[f(\boldsymbol\beta)\mid D_{1:t}]$ as $M\to\infty$, with fixed-horizon error $O_p(M^{-1/2})$.
\end{proposition}

\begin{proposition}[Bayesian calibration under correct specification]
If the model is correctly specified and posterior inference is exact, then posterior predictive verdict probabilities are calibrated in the Bayesian sense: conditional on a posterior predictive event probability, the posterior expected frequency of that event equals the probability itself.
\end{proposition}
This is not a finite-sample distribution-free coverage guarantee, and it does not automatically apply to the MAP-profile/hard-assignment surrogate. When exact coverage is required, conformal methods can be layered on top of BayesJudge \citep{angelopoulos2021gentle,sheng2025analyzing}.

\subsection{Antisymmetric bias-covariate extension}
\label{app: extension}
The scalar model can be extended by replacing $\pm\beta_j$ with $b_j^\top u_{tj}^{(o)}$, where $u_{tj}^{(BA)}=-u_{tj}^{(AB)}$. Then
\begin{align}
  \xi_{tj,A}^{(o)}=\theta_t+b_j^\top u_{tj}^{(o)},
  \qquad
  \xi_{tj,T}^{(o)}=\tau_t,
  \qquad
  \xi_{tj,B}^{(o)}=-\theta_t-b_j^\top u_{tj}^{(o)}.
\end{align}
In principle, any bias source admitting an antisymmetric encoding, i.e., a feature whose sign flips under AB/BA swapping such as length-difference or formatting-difference effects, can be modeled in exactly the same way, with position bias ($u_{tj}^{(\mathrm{AB})} = 1$) as the simplest special case; the identifiability analysis and empirical validation of such extensions are left to future work.

\section{Hyperparameter Setting And Sensitivity Analysis}
\label{app:sensitive}

\subsection{Experimental Setting and Details}
\label{app:loho}
\paragraph{Leave-One-Human-Out (LOHO) validation.} For each held-out rater $r$, we infer the item posterior using all remaining human labels $\{y_{tr'}^H\}_{r' \neq r}$ and all LLM swap judgments. The predictive distribution for $r$ is obtained by marginalizing over the posterior item verdict and the inferred confusion structure:
\[
p\bigl(y_{tr}^H \mid \text{rest}\bigr) = \sum_{k} \hat{\pi}_{t,k} \, \hat{C}_r^{(t-1)}[k, \cdot].
\]
Class priors are implicitly uniform under the BTD prior. We evaluate against the held-out rater's observed labels using three standard metrics:

\textbf{Accuracy (Acc).} The predicted label is the argmax of the marginalized predictive distribution:
\[
\hat{y}_{tr}^H = \arg\max_{c \in \mathcal{K}} \, p\bigl(y_{tr}^H = c \mid \text{rest}\bigr).
\]
Accuracy is the fraction of correct predictions: $\text{Acc} = \frac{1}{T_{\text{test}}} \sum_{t} \mathbbm{1}[\hat{y}_{tr}^H = y_{tr}^H]$.

\textbf{Negative Log-Likelihood (NLL).} The per-sample NLL is:
\[
\text{NLL}_t = -\log p\bigl(y_{tr}^H \mid \text{rest}\bigr),
\]
averaged over the test set.

\textbf{Expected Calibration Error (ECE).} We partition predictions into $B=10$ equal-mass bins by confidence (the max predictive probability). Let $B_b$ denote the set of samples in bin $b$, with average confidence $\bar{p}_b$ and average accuracy $\hat{a}_b$. Then:
\[
\text{ECE} = \sum_{b=1}^{10} \frac{|B_b|}{N} \bigl| \bar{p}_b - \hat{a}_b \bigr|.
\]

The hyperparameter setting in BayesJudge is shown in Tabel \ref{tab:hyperparams}.
\begin{table}[H]
\centering
\caption{Core hyperparameters of BayesJudge.}
\label{tab:hyperparams}
\small
\begin{tabular}{@{}lll@{}}
\toprule
\textbf{Parameter} & \textbf{Default} & \textbf{Description} \\
\midrule
$M$ & 300 & Number of SMC particles \\
$\kappa$ & 0.5 & ESS resampling threshold factor ($\text{threshold} = \kappa M$) \\
$\epsilon_0$ & 0.5 & Initial MCMC move step size (actual step size $\epsilon_t = \epsilon_0 / \sqrt{t}$) \\
\midrule
$\lambda_{\mathrm{diag}}$ & 1.0 & Diagonal concentration of the Dirichlet prior for human confusion matrices \\
$\lambda_{\mathrm{off}}$ & 0.05 & Off-diagonal concentration; zero values in ablations denote heuristic\\ 
&& pseudo-count initialization with a numerical floor \\
$\sigma_{\beta}$ & 1.0 & Prior standard deviation of LLM position bias $\beta_j$ \\
$\mu_{0}$ & $[0.0,\, 0.0]^{\top}$ & Prior mean of local item parameters $\eta_t = (\theta_t, \tau_t)$ \\
$\Sigma_{0}$ & $I$ & Prior covariance matrix of local item parameters $\eta_t$ \\
\bottomrule
\end{tabular}
\end{table}

\subsection{Human Confusion Matrix Recovery And Prior Analysis}
\label{app:human_confusion_analysis}
\subsubsection{Different Initialization.}
\label{app:init_analysis}
Table~\ref{tab:sensitivity} summarizes the reported sensitivity study for the human confusion priors in synthetic dataset. When the initialization is diagonally dominant, recovery is stable across prior choices; the coefficient of variation of the final error is only $5.76\%$ for the expert and $3.11\%$ for the novice. When the prior is strongly off-diagonal, the expert rater becomes substantially more sensitive, with error variation rising to $45.88\%$. This asymmetry is interpretable: an expert-like ground-truth rater is structurally mismatched with off-diagonal priors, whereas a novice-like rater is closer to diffuse confusion.

\begin{table}[H]
\centering
\small
\caption{Sensitivity of recovered human confusion matrices to initialization. Frobenius errors are from the completed synthetic study.}
\label{tab:sensitivity}
\begin{tabular}{lcccc}
\toprule
\multirow{2}{*}{Initialization $(\lambda,\gamma)$} & \multicolumn{2}{c}{Diagonal-dominant} & \multicolumn{2}{c}{Off-diagonal-dominant} \\
\cmidrule(lr){2-3}\cmidrule(lr){4-5}
& Expert & Novice & Expert & Novice \\
\midrule
$(0.5,0.0)/(0.1,1.0)$ & 0.0693 & 0.0538 & 0.0833 & 0.0535 \\
$(1.0,0.0)/(0.2,0.5)$ & 0.0653 & 0.0543 & 0.0694 & 0.0518 \\
$(2.0,0.0)/(0.5,1.0)$ & 0.0641 & 0.0549 & 0.0788 & 0.0506 \\
$(5.0,0.0)/(0.5,2.0)$ & 0.0640 & 0.0588 & 0.1150 & 0.0761 \\
$(1.0,0.1)/(1.0,1.0)$ & 0.0661 & 0.0541 & 0.0741 & 0.0488 \\
$(1.0,0.5)/(1.0,2.0)$ & 0.0694 & 0.0532 & 0.1005 & 0.0564 \\
$(2.5,0.5)/(1.0,5.0)$ & 0.0709 & 0.0551 & 0.1932 & 0.0379 \\
$(5.0,1.0)/(2.0,5.0)$ & 0.0753 & 0.0549 & 0.1971 & 0.0314 \\
\midrule
Coefficient of variation & 5.76\% & 3.11\% & 45.88\% & 26.05\% \\
\bottomrule
\end{tabular}
\end{table}

\begin{figure}[H]
\centering
\includegraphics[width=0.95\linewidth]{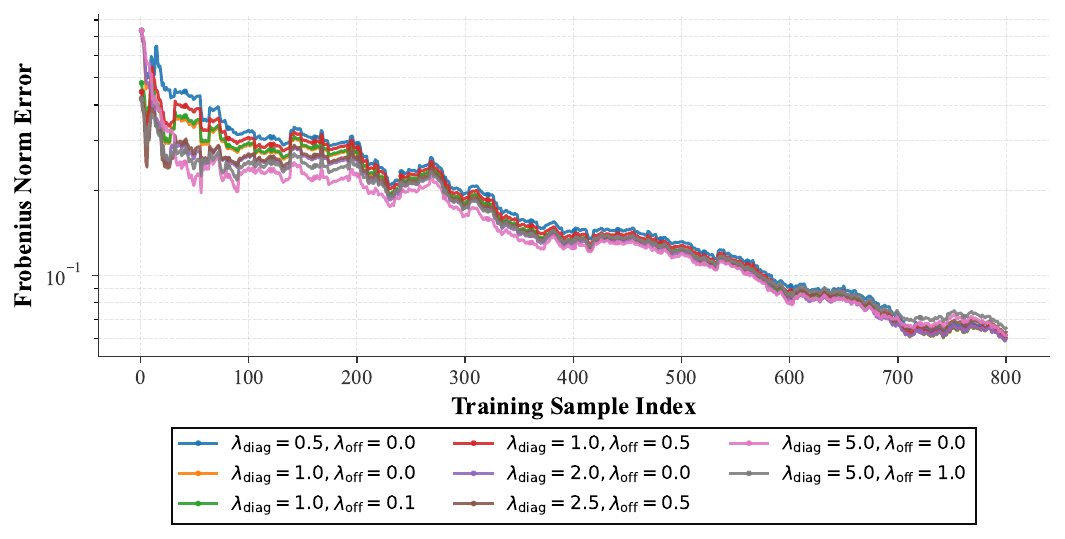}
\caption{Impact of initialization strategies on human confusion matrix convergence under diagonal dominance. Plotted in log scale.}
\label{appfig:init_diag_fnorm}
\end{figure}

\begin{figure}[H]
\centering
\includegraphics[width=0.95\linewidth]{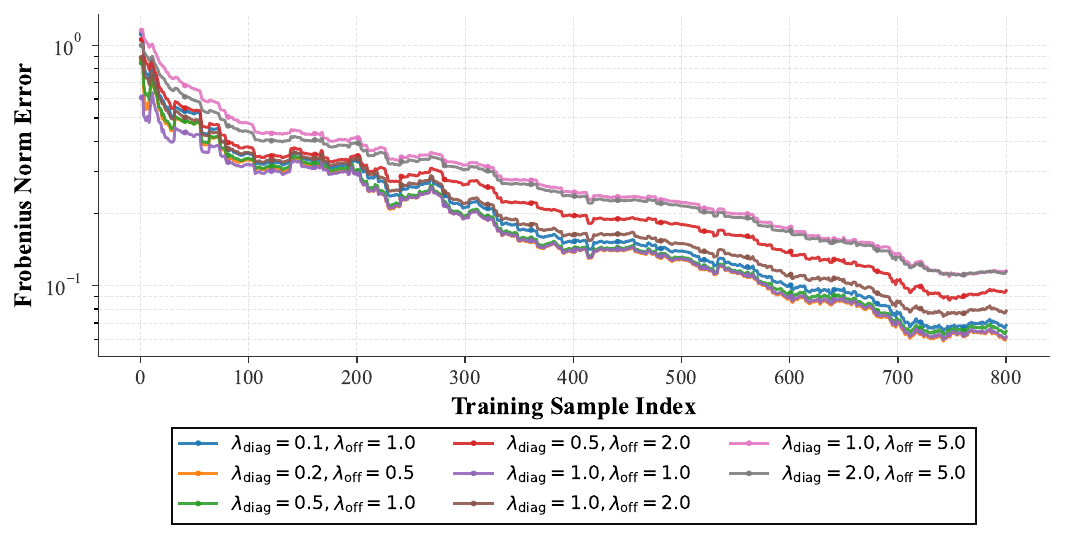}
\caption{Impact of initialization strategies on human confusion matrix convergence under off-diagonal dominance.  Plotted in log scale.}
\label{appfig:init_offdiag_fnorm}
\end{figure}

We also visualize the impact of Dirichlet prior initialization on the recovery accuracy of human confusion matrices as shown in Figure \ref{appfig:init_diag_fnorm}, \ref{appfig:init_offdiag_fnorm}. Two experimental regimes are considered. In the first (\texttt{init\_ablation}), the off-diagonal prior concentration is fixed at a low level ($\lambda_{\text{off}} \in [0.0, 1.0]$) while the diagonal concentration is systematically varied ($\lambda_{\text{diag}} \in [0.5, 5.0]$). In the second (\texttt{init\_ablation\_offdiag}), the off-diagonal concentration is the primary variable ($\lambda_{\text{off}} \in [0.5, 5.0]$).

\textbf{Finding 1: Diagonal prior concentration is highly robust.} Under the first regime, all eight $(\lambda_{\text{diag}}, \lambda_{\text{off}})$ configurations yield tightly clustered final Frobenius errors (0.0595--0.0651), with relative variation below 9\%. This indicates that, provided the off-diagonal prior remains weak ($\lambda_{\text{off}} \leq 1.0$), the online Dirichlet cumulative update mechanism effectively overrides differences in diagonal prior concentration and converges to similar posterior estimates. The underlying reason is that the synthetic human confusion matrices exhibit strong diagonal dominance (expert raters with diagonal elements $\geq 0.90$), so the strong likelihood signal from observed data rapidly overwhelms prior discrepancies.

\textbf{Finding 2: Excessive off-diagonal prior induces severe estimation bias.} The second regime reveals a markedly different picture: as $\lambda_{\text{off}}$ increases from 0.5 to 5.0, the final error deteriorates from 0.0606 to 0.1155, an amplification of nearly 90\%. Notably, the configurations $(\lambda_{\text{diag}}=1.0, \lambda_{\text{off}}=5.0)$ and $(2.0, 5.0)$ perform almost identically and both very poorly. This demonstrates that once the off-diagonal prior becomes dominant, further increasing the diagonal concentration cannot rescue estimation quality. The Dirichlet prior then artificially assigns excessive prior probability to mislabeling events, creating a structural mismatch with the true data-generating process (low mislabeling rate) and forcing the posterior to allocate undue mass in off-diagonal regions.

\textbf{Practical Recommendation.} These findings suggest adopting a conservative prior strategy with weak positive off-diagonal and moderate diagonal concentrations (e.g., $\lambda_{\text{diag}} \in [1.0, 2.0]$, $\lambda_{\text{off}} \in [10^{-3}, 0.5]$) in production deployments. The off-diagonal prior should not exceed 1.0, as values beyond this threshold may introduce systematic biases that data likelihood cannot adequately correct. This aligns with standard practice in the Dawid--Skene literature, where rater priors are typically modeled as slight perturbations of the identity matrix rather than uniform or inversely skewed matrices.

\subsubsection{Different Numbers of LLM Evaluators}
\begin{figure}[H]
\centering
\includegraphics[width=1.0\linewidth]{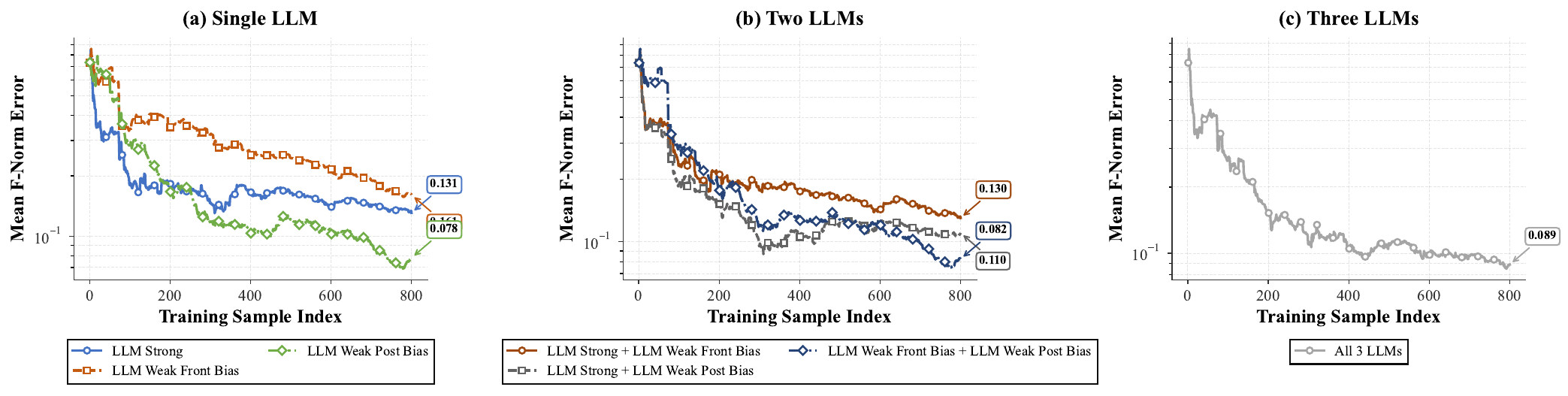}
\caption{Comparison of Human Confusion Matrix Convergence Curves Under Different LLM Configurations}
\label{appfig:llm_count_f_norm}
\end{figure}

We further investigate the impact of LLM evaluator count on human confusion matrix recovery. Specifically, we simulate three distinct LLM profiles: a strong LLM (no position bias), a weak LLM with front-position bias, and a weak LLM with back-position bias. Figure~\ref{appfig:llm_count_f_norm} shows the online convergence of mean Frobenius error in $\hat{\beta}_j$ as the number and composition of LLM judges vary. Under single-judge configurations (panel~(a)), the post-position-biased judge (\textit{weak-post}) achieves the lowest asymptotic error ($0.0782$), whereas the front-position-biased judge (\textit{weak-front}) suffers the highest ($0.1608$). The nominally unbiased \textit{strong} judge sits in between ($0.1309$), suggesting that a judge with a \emph{consistent, identifiable} antisymmetric bias is easier to correct for than one whose bias is small but not structurally pinned down by the swap likelihood.

The most striking result appears in panel~(b): a two-judge panel composed of \textit{weak-front} and \textit{weak-post} judges attains a final error of only $0.0824$, outperforming not only every single-judge baseline but also the mixed panels (\textit{strong}+\textit{weak-front} at $0.1302$, \textit{strong}+\textit{weak-post} at $0.1102$). Because the two weak judges carry biases of opposite sign, the swap likelihood’s antisymmetric structure allows BayesJudge to \emph{cancel} their presentation effects while jointly constraining the item preference $\theta_t$. This yields tighter posterior contraction than relying on a solitary, nominally high-quality judge.

Adding a third judge (panel~(c)) produces a final error of $0.0888$, slightly \emph{worse} than the optimal two-judge pair. The marginal gain of the extra judge is evidently outweighed by the increased dimensionality of the global parameter space and the finite particle budget of the SMC sampler. Taken together, these results suggest a practical protocol design principle: when multiple LLM judges are available, prefer a \emph{small, complementary} panel with symmetrically opposing biases over a larger or uniformly ``strong'' ensemble; BayesJudge’s swap likelihood turns their opposing distortions into mutually identifying information rather than compounded noise.

\subsection{Robustness to Adversarial Raters}
\label{app:adversarial}

We test whether an adversarial annotator can corrupt the joint estimates. We
augment the synthetic setup of Section~5.2 (1{,}000 samples, 800/200 split,
300 particles, 2 regular human annotators, 2 LLM judges) with one adversarial
rater who outputs $A$ on every item regardless of the true verdict.

\begin{table}[h]
  \caption{Effect of an always-$A$ adversarial rater (synthetic).}
  \label{tab:adversarial}
  \centering
  \small
  \begin{tabular}{lcc}
    \toprule
    Metric & No adversary & With adversary \\
    \midrule
    Adversary row entropy & --- & $\approx 0.00$--$0.04$ \\
    Normal-expert Frob.\ error & 0.065 & 0.048 \\
    Mean item posterior entropy & 0.8176 & 0.8177 \\
    \bottomrule
  \end{tabular}
\end{table}

The Dirichlet posterior isolates the adversarial pattern: the adversary's
confusion matrix converges to $[1.0, 0.0, 0.0]$ in all three rows, with row
entropy near zero (fully deterministic behavior rather than noise). This
behavior is absorbed into the rater-specific $C_r$ without leaking into the
item posterior: mean $\pi_t$ entropy changes by only $+0.0001$, and the
normal expert's confusion-matrix error remains stable ($0.065 \to 0.048$).
Adversarial behavior is thus quarantined in rater-specific parameters rather
than misinterpreted as item ambiguity or injected into global estimates.

\section{SMC Particle Health Diagnostics}
\label{app:ess}
\begin{figure}[H]
\centering
\includegraphics[width=1.0\linewidth]{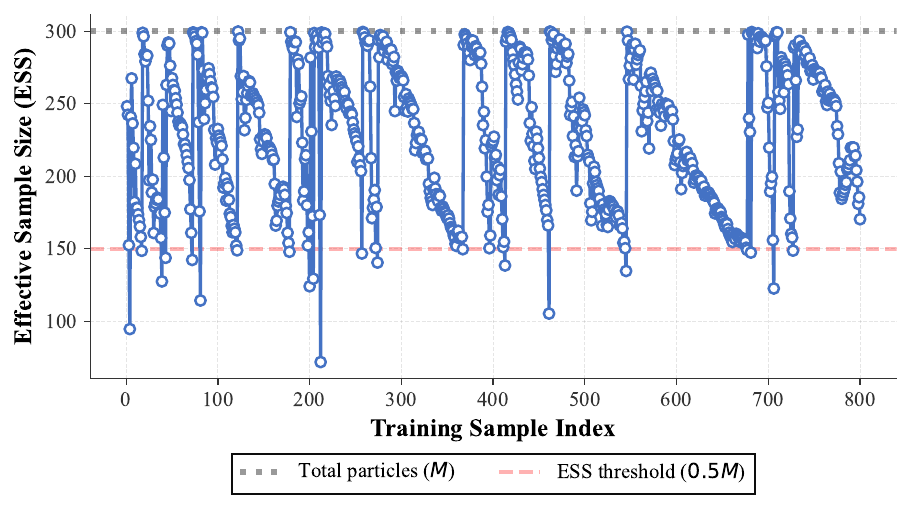}
\caption{ESS trajectory of the SMC particle cloud on the SummEval stream with GPT-4o-mini ($M=300$, $\kappa=0.5$). The characteristic saw-tooth pattern indicates healthy online particle filtering: after each systematic resampling, the ESS is restored to near $M$ (grey dashed line), then gradually decays until the threshold $\kappa M=150$ (red dashed line) triggers the next resample-move cycle, confirming stable inference in the real-world regime.}
\label{appfig:ess}
\end{figure}

Figure~\ref{appfig:ess} traces the effective sample size (ESS) of the SMC particle cloud over the SummEval training stream ($M=300$ particles, resampling threshold $\kappa M = 150$). The characteristic saw-tooth pattern is the hallmark of healthy online particle filtering: after each systematic resampling, the ESS is restored to near $M$ (the grey dashed line at 300), then gradually decays as sequential importance weights accumulate heterogeneity across items. Whenever the ESS drops below the threshold of $0.5M$ (red dashed line), resampling is triggered, producing the sharp vertical recoveries observed throughout the stream.

Two features are worth noting. First, the inter-resampling interval varies between roughly 20 and 60 items, indicating that the rate of particle degeneracy is data-dependent: items with conflicting human and LLM evidence induce faster weight skew, whereas homogeneous items allow the cloud to remain stable longer. Second, occasional deep troughs—most prominently near sample 200, where the ESS briefly falls to approximately 75—coincide with high-ambiguity items for which the swap likelihood and human channel place mass on different verdicts. These troughs never trigger catastrophic collapse because the resample-move rejuvenation (Algorithm~1, line 38) restores diversity before the next update.

Overall, the ESS trajectory confirms that the Rao--Blackwellized assumed-density SMC implementation remains numerically stable: the particle cloud spends the majority of its lifetime above the $0.5M$ safety threshold, and the periodic resampling schedule keeps the Monte Carlo variance of global bias estimates $\hat{\beta}_j$ under control without excessive computational overhead.

\section{Computational Efficiency}
\label{app:compute}
\subsection{Hard vs.\ Soft Assignment for Human Confusion Updates}
Table~\ref{tab:hard-soft-summary} contrasts the two update strategies. Soft assignment incurs a $22$--$38\%$ runtime overhead while delivering substantially worse recovery: the human confusion Frobenius error deteriorates from $0.060$ (hard) to $0.393$ (soft), a $6.6\times$ gap. LLM bias RMSE also rises by $\approx 9\%$. Resampling frequency is unaffected, confirming that the overhead is localized to the Dirichlet update rather than SMC degeneracy.

\begin{table}[h]
\centering
\caption{Hard vs.\ soft assignment: summary comparison (synthetic stream, $T=800$).}
\label{tab:hard-soft-summary}
\footnotesize
\setlength{\tabcolsep}{4pt}
\begin{tabular}{@{}lccc@{}}
\toprule
\textbf{Metric} & \textbf{Hard} & \textbf{Soft} & \textbf{Gap} \\
\midrule
Per-item time, $M=50$ (ms)  & 145.20 & 177.17 & Soft $+22\%$ slower \\
Per-item time, $M=1000$ (ms)& 2058.75& 2851.48& Soft $+38\%$ slower \\
Human $C$ F-norm (final)    & 0.0598 & 0.3931 & Soft $6.6\times$ worse \\
Resample events             & 18--25 & 18--23 & Comparable \\
\bottomrule
\end{tabular}
\end{table}

\paragraph{Scaling with particle budget.}
Table~\ref{tab:runtime-hard-soft} breaks down runtime by $M$. Both variants scale roughly linearly, but the soft-update overhead widens from $14\%$ at $M=100$ to a peak of $42.8\%$ at $M=500$. The extra cost stems from distributing each pseudo-count across all three Dirichlet rows under soft assignment, whereas hard update commits to a single cell.

\begin{table}[h]
\centering
\caption{Runtime scaling: hard vs.\ soft assignment across particle budgets.}
\label{tab:runtime-hard-soft}
\footnotesize
\setlength{\tabcolsep}{3pt}
\begin{tabular}{@{}lccccc@{}}
\toprule
& \multicolumn{2}{c}{\textbf{Hard Update}} & \multicolumn{2}{c}{\textbf{Soft Update}} & \textbf{Relative} \\
\cmidrule(lr){2-3}\cmidrule(lr){4-5}
$M$ & Total (s) & Per-item (ms) & Total (s) & Per-item (ms) & Slowdown \\
\midrule
50   & 116.16 & 145.20 & 141.74 & 177.17 & $+22.0\%$ \\
100  & 222.50 & 278.12 & 253.71 & 317.14 & $+14.0\%$ \\
200  & 426.79 & 533.48 & 558.75 & 698.43 & $+30.9\%$ \\
300  & 613.43 & 766.79 & 853.68 & 1067.11& $+39.2\%$ \\
500  & 913.97 & 1142.46& 1305.30& 1631.63& $+42.8\%$ \\
1000 & 1647.00& 2058.75& 2281.18& 2851.48& $+38.5\%$ \\
\bottomrule
\end{tabular}
\end{table}

\paragraph{Discussion.}
The accuracy degradation under soft assignment is an empirical phenomenon rather than a theoretical necessity. In the online regime, early particle weights are volatile; soft responsibilities smear Dirichlet counts across multiple rows of $C_r$, diluting the diagonal signal. Hard assignment acts as a winner-take-all regularizer that aggressively commits the posterior to the most likely confusion pattern. All main-text experiments use the hard-MAP assignment update in Equation~\eqref{eq:hard_assignment}, because it was more stable in the online assumed-density regime. The soft responsibility update is retained as a Rao--Blackwellized ablation rather than the default implementation.

\subsection{Exact-Inference Surrogates: Accuracy--Cost Comparison}
\label{app:inference_comparison}

The streaming implementation replaces the exact local marginal likelihood
$m_t(\Phi)$ with a MAP profile (Eq.~\ref{eq:profile_potential}) and replaces marginalization
of the latent verdict with a hard assignment (Eq.~\ref{eq:hard_assignment}). To quantify what these
approximations cost, we compare the implemented surrogate against two
stronger inference schemes on synthetic data ($M{=}300$ particles, 800
training and 200 test samples): a Laplace-corrected potential (Eq.~\ref{eq:laplace_potential}) and a
random-walk Metropolis--Hastings chain targeting the exact local posterior.

\begin{table}[h]
  \caption{Inference-scheme comparison on synthetic data.}
  \label{tab:inference_comparison}
  \centering
  \small
  \begin{tabular}{lccc}
    \toprule
    & MAP + hard update (ours) & Laplace & MCMC \\
    \midrule
    Train time & 180\,s & 347\,s & 1{,}803\,s \\
    Per-sample latency & 230\,ms & 434\,ms & 2{,}253\,ms \\
    $C_r$ Frob.\ error --- expert & 0.0653 & 0.0702 & 0.1045 \\
    $C_r$ Frob.\ error --- novice & 0.0543 & 0.0799 & 0.0428 \\
    Test log-likelihood & $-5.336$ & $-5.401$ & $-5.396$ \\
    Test perplexity & 210.0 & 221.5 & 220.4 \\
    \bottomrule
  \end{tabular}
\end{table}

The implemented surrogate attains the lowest expert-matrix error and the best
test log-likelihood while being roughly $10\times$ faster than MCMC; the MCMC
reference achieves the lowest novice-matrix error. These results support the
surrogate's computational utility in this controlled setting; they do not
establish posterior equivalence, and the exact-posterior calibration
statements of Appendix~C.10 do not automatically transfer to the surrogate.
\section{Algorithm Pseudocode}
\label{app:alg}
We present the complete algorithmic pseudocode for the streaming approximation in Algorithms~\ref{alg:smc},~\ref{alg:maplocal}, and~\ref{alg:predll}. The input LLM labels are assumed to have been canonicalized by the mapping in Section~\ref{sec:model}.

\begin{algorithm}[H]
\caption{Online MAP-Assisted Joint SMC Inference}
\label{alg:smc}
\begin{algorithmic}[1]
\Require Sample stream $\{s_t\}_{t=1}^T$, particle number $M$, ESS threshold factor $\gamma$, move step $\epsilon_0$, priors $(\boldsymbol{\mu}_\eta,\boldsymbol{\Sigma}_\eta,\lambda_{\rm diag},\lambda_{\rm off},\sigma_\beta)$, assignment mode $\textsc{Mode}\in\{\textsc{Hard},\textsc{Soft}\}$
\Ensure Joint particles $\{\boldsymbol\beta^{(m)},w^{(m)}\}_{m=1}^M$; Dirichlet statistics $\{\boldsymbol\alpha_r\}$
\State Initialize an empty judge set $\mathcal J_{\rm seen}$ and rater set $\mathcal R_{\rm seen}$
\State Set $w^{(m)}\gets 1/M$ for $m=1,\dots,M$
\For{each sample $s_t=(\mathbf y^H,\mathbf y^L)$}
    \For{each newly observed judge $j\in J_t\setminus\mathcal J_{\rm seen}$}
        \For{$m=1$ to $M$}
            \State Sample $\beta_j^{(m)}\sim\mathcal N(0,\sigma_\beta^2)$ and append it to joint particle $\boldsymbol\beta^{(m)}$
        \EndFor
        \State $\mathcal J_{\rm seen}\gets\mathcal J_{\rm seen}\cup\{j\}$
    \EndFor
    \For{each newly observed rater $r\in R_t\setminus\mathcal R_{\rm seen}$}
        \State $\alpha_{r,k,\ell}\gets\lambda_{\rm off}+ (\lambda_{\rm diag}-\lambda_{\rm off})\mathbb I[k=\ell]$ for all $k,\ell\in\mathcal K$
        \State $\mathcal R_{\rm seen}\gets\mathcal R_{\rm seen}\cup\{r\}$
    \EndFor
    \For{each $r\in R_t$}
        \State $\widehat C_r[k,\ell]\gets \alpha_{r,k,\ell}/\sum_{\ell'}\alpha_{r,k,\ell'}$
    \EndFor
    \For{$m=1$ to $M$}
        \State $\widehat\eta_t^{(m)}\gets\textsc{MapLocal}(\mathbf y^H,\mathbf y^L,\boldsymbol\beta^{(m)},\{\widehat C_r\}_{r\in R_t})$
        \State $\widehat m_t^{(m)}\gets\exp\{\mathcal L_t^{(m)}(\widehat\eta_t^{(m)})\}$ \Comment{profile potential; Laplace optional}
        \State $\log \widetilde w^{(m)}\gets \log w^{(m)}+\log \widehat m_t^{(m)}$
    \EndFor
    \State $\mathbf w\gets\textsc{LogNormalize}(\{\log \widetilde w^{(m)}\}_{m=1}^M)$
    \State $\ESS\gets1/\sum_{m=1}^M(w^{(m)})^2$
    \If{$\ESS<\gamma M$}
        \State $\{\iota_m\}_{m=1}^M\gets\textsc{SysResample}(\mathbf w)$
        \State $\boldsymbol\beta^{(m)}\gets\boldsymbol\beta^{(\iota_m)}$, $\widehat\eta_t^{(m)}\gets\widehat\eta_t^{(\iota_m)}$ for all $m$
        \State $w^{(m)}\gets1/M$ for all $m$
    \EndIf
    \State $\epsilon_t\gets\epsilon_0/\sqrt t$
    \State $\{\boldsymbol\beta^{(m)}\}_{m=1}^M\gets\textsc{JointMove}(\{\boldsymbol\beta^{(m)}\}_{m=1}^M,\mathcal B_t,\sigma_\beta,\epsilon_t)$
    \State $\bar\pi_{t,k}\gets\sum_{m=1}^M w^{(m)}g_k(\widehat\eta_t^{(m)})$ for each $k\in\mathcal K$ 
    \If{$\textsc{Mode}=\textsc{Hard}$}
        \State $k^\star\gets\arg\max_{h}\bar\pi_{t,h}$, \quad $q_{t,k}\gets\mathbb I[k=k^\star]$
    \Else
        \State $q_{t,k}\gets\bar\pi_{t,k}$
    \EndIf
    \For{each $r\in R_t$}
        \State $o_r\gets\textsc{Idx}(y_{tr}^H)$
        \For{each $k\in\mathcal K$}
            \State $\alpha_{r,k,o_r}\gets\alpha_{r,k,o_r}+q_{t,k}$
        \EndFor
    \EndFor
\EndFor
\end{algorithmic}
\end{algorithm}

\begin{algorithm}[H]
\caption{Local MAP Estimation (\textsc{MapLocal})}
\label{alg:maplocal}
\begin{algorithmic}[1]
\Require Human labels $\mathbf y^H$, canonicalized LLM labels $\mathbf y^L$, joint LLM-bias particle $\boldsymbol\beta$, confusion matrices $\{\widehat C_r\}$
\Ensure Local parameter estimate $\widehat\eta_t=(\widehat\theta_t,\widehat\tau_t)$
\State $H(\eta)\gets\prod_{r\in R_t}\sum_{k\in\mathcal K}g_k(\eta)\,\widehat C_r[k,\textsc{Idx}(y_{tr}^H)]$ \Comment{per-rater marginal, Eq. (5)}
\State $\mathcal L_L(\eta)\gets\sum_{j\in J_t}\left[\log p(y_{tj}^{L,(AB)}\mid\eta,\beta_j)+\log p(y_{tj}^{L,(BA)}\mid\eta,\beta_j)\right]$
\State $\mathcal L(\eta)\gets -\tfrac12(\eta-\boldsymbol\mu_\eta)^\top\boldsymbol\Sigma_\eta^{-1}(\eta-\boldsymbol\mu_\eta)+\log H(\eta)+\mathcal L_L(\eta)$
\State $\widehat\eta_t\gets\arg\max_\eta\mathcal L(\eta)$ \Comment{solved by L-BFGS-B or short MH}
\State \Return $\widehat\eta_t$
\end{algorithmic}
\end{algorithm}

\begin{algorithm}[H]
\caption{Predictive Log-Likelihood Evaluation}
\label{alg:predll}
\begin{algorithmic}[1]
\Require New labels $(\mathbf y^H,\mathbf y^L)$, joint posterior particles $\{\boldsymbol\beta^{(m)},w^{(m)}\}_{m=1}^M$, posterior mean confusion matrices $\{\widehat C_r\}$, Monte Carlo size $N_{\rm mc}$
\Ensure Predictive marginal log-likelihood $\log\widehat p(\mathbf y^H,\mathbf y^L\mid D_{1:t})$
\For{$i=1$ to $N_{\rm mc}$}
    \State Sample $\eta_i\sim\mathcal N(\boldsymbol\mu_\eta,\boldsymbol\Sigma_\eta)$
    \State Sample a joint particle index $m_i\sim\Cat(w^{(1)},\dots,w^{(M)})$ and set $\boldsymbol\beta_i\gets\boldsymbol\beta^{(m_i)}$
    \State $H_i\gets\prod_{r\in R_{\rm new}}\sum_{k\in\mathcal K}g_k(\eta_i)\,\widehat C_r[k,y_r^H]$ 
    \State $\ell_i\gets\log H_i+\sum_{j\in J_{\rm new}}\log L_j^{\rm swap}(\eta_i,\beta_{ij})$
\EndFor
\State \Return $\textsc{LogSumExp}(\{\ell_i\}_{i=1}^{N_{\rm mc}})-\log N_{\rm mc}$
\end{algorithmic}
\end{algorithm}
\section{Prompt}
\label{app:prompt}
The following is the prompt template used for pairwise evaluation by LLM in our experiments.

\begin{tcolorbox}[colback=gray!5!white,colframe=gray!75!black,title={\textbf{Pairwise Evaluation Prompt}},fonttitle=\bfseries,boxrule=0.5pt,arc=2pt]
\ttfamily\small\raggedright
\setlength{\parindent}{0pt}
\setlength{\parskip}{0.6em}

Please act as an impartial judge and evaluate the quality of the summaries provided by two AI assistants to the source text displayed below.

Please only output a single letter to indicate your choice.

\hangindent=1.5em \hangafter=1
- Output A if Assistant A's summary is better.

\hangindent=1.5em \hangafter=1
- Output B if Assistant B's summary is better.

\hangindent=1.5em \hangafter=1
- Output C if they are tied.

\textbf{Your Output Should Be Only: A or B or C}

\vspace{0.5em}
\noindent\rule{\linewidth}{0.4pt}
\vspace{0.5em}

\noindent\textbf{\#\# Source Text}\\
\{text\}

\noindent\textbf{\#\# Assistant A}\\
\{res\_a\}

\noindent\textbf{\#\# Assistant B}\\
\{res\_b\}
\end{tcolorbox}

\noindent\textbf{Note:} The placeholders \texttt{\{text\}}, \texttt{\{res\_a\}}, and \texttt{\{res\_b\}} are dynamically filled with the source document and the two candidate summaries at training and inference time. The LLM outputs raw options A, B, or C, where C is mapped to the canonical tie label T. For BA prompts, raw prompt-position labels are mapped back to canonical response coordinates before applying the likelihood: raw A becomes canonical B, raw B becomes canonical A, and raw C remains T. Thus the mathematical labels A and B always refer to the original candidate identities, not prompt positions.

\section{Supplementary Analysis}
\label{app:supp}
\subsection{Ablation Study on SummEval}
\paragraph{Held-out predictive log-likelihood (PLL).}
We evaluate joint model calibration via the marginal predictive density on held-out items. For a test sample $Y_{\text{test}}=(Y^H,Y^L)$, we approximate the intractable integral over parameters by Monte Carlo:
\begin{itemize}[leftmargin=*,nosep]
    \item Sample item parameters from the prior: $\eta_i \sim \mathcal{N}(\mu_\eta, \Sigma_\eta)$, $i=1,\dots,N$ ($N=1000$);
    \item Sample judge biases from the SMC posterior: $\beta_j^{(i)} \sim \text{Categorical}\big(\{\beta_j^{(m)}, w_j^{(m)}\}_{m=1}^M\big)$.
\end{itemize}
The per-sample joint log-likelihood decomposes across channels:
\begin{equation}
\ell_i = \underbrace{\sum_{r\in\mathcal{R}_{\text{test}}} \log \big[(\hat{C}_r^\top g(\eta_i))_{y_r^H}\big]}_{\text{human channel}} + \underbrace{\sum_{j\in\mathcal{J}_{\text{test}}} \log \mathcal{L}_j^{\text{swap}}(\eta_i, \beta_j^{(i)})}_{\text{LLM swap channel}}.
\end{equation}
The PLL is then computed by the numerically stable LogSumExp estimator:
\begin{equation}
\widehat{\text{PLL}} = \text{LogSumExp}\big(\{\ell_i\}_{i=1}^N\big) - \log N
\;=\; \log\!\left(\frac{1}{N}\sum_{i=1}^N \exp(\ell_i)\right).
\end{equation}
This quantity integrates out parameter uncertainty: $\eta$ is drawn from its generative prior (reflecting the assumed item-quality distribution), while $\beta_j$ is drawn from its streaming SMC posterior (reflecting accumulated evidence about judge-specific bias). Higher PLL (closer to zero) indicates better joint calibration of the item verdict, rater confusion, and LLM position-bias estimates. A difference of $\Delta$ nats corresponds to a multiplicative factor of $\exp(\Delta)$ in average predictive probability.

Ablation results on SummEval dataset are shown in Table \ref{tab:ablation_summeval}.

\begin{table}[h]
\centering
\caption{Ablation on held-out predictive log likelihood (SummEval). Higher is better.}
\label{tab:ablation_summeval}
\resizebox{0.6\columnwidth}{!}{%
\begin{tabular}{lc}
\toprule
Model variant & Log likelihood \\
\midrule
BayesJudge & $\mathbf{-9.7766}$ \\
w/o human rater update ($C_r=I$) & $-9.8485$ \\
w/o LLM bias update ($\beta_j=0$) & $-9.9038$ \\
w/o both updates & $-10.0507$ \\
\midrule
Improvement (full vs. w/o both) & $+0.2741$ \\
\bottomrule
\end{tabular}%
}
\end{table}

\subsection{Rater Sparsity and Expert--Crowd Separation}
\label{app:rater-sparsity}

SummEval contains an imbalanced rater pool ($3$ experts vs.\ $5$ crowdworkers). We investigate how the \emph{number} and \emph{composition} of available human raters affect the unsupervised identification of their confusion signatures. Because no ground-truth $C_r$ exists on real data, we assess recovery via (i) LOHO predictive accuracy, which is sensitive to misestimated confusion, and (ii) the posterior structural gap between expert and crowd confusion matrices.

\paragraph{Experimental setup.}
From the full panel, we draw stratified subsamples while preserving the $3\!:\!5$ expert--crowd ratio (rounded). BayesJudge is run with GPT-4o-mini fixed as the LLM judge. Table~\ref{tab:rater-sparsity-setup} details the panel configurations.

\begin{table}[h]
\centering
\caption{Rater-panel configurations for the sparsity study (SummEval).}
\label{tab:rater-sparsity-setup}
\footnotesize
\setlength{\tabcolsep}{4pt}
\begin{tabular}{@{}lcccc@{}}
\toprule
\textbf{Total raters} $n$ & \textbf{Experts} $n_{\text{exp}}$ & \textbf{Crowd} $n_{\text{crowd}}$ & \textbf{Expert ratio} & \textbf{Stratification} \\
\midrule
2  & 1 & 1 & 50.0\%  & Minimum viable panel (both types present) \\
4  & 1 & 3 & 25.0\%  & Crowd-dominated, mimicking low-budget reality \\
6  & 2 & 4 & 33.3\%  & Near-original $3\!:\!5$ proportion \\
8  & 3 & 5 & 37.5\%  & Full panel (reference) \\
\bottomrule
\end{tabular}
\end{table}

\paragraph{Evaluation protocol.}
For each panel configuration:
\begin{itemize}[leftmargin=*,nosep]
    \item \textbf{Training:} Run BayesJudge on the SummEval training stream ($80\%$) to obtain per-rater posterior confusion matrices $\hat{C}_r$ and LLM bias $\hat{\beta}_j$.
    \item \textbf{Test (LOHO):} On the held-out test set ($20\%$), withhold one rater $r^*$ and predict their labels from the remaining human observations plus LLM swap judgments. Report accuracy stratified by rater type (expert vs.\ crowd).
    \item \textbf{Structural gap:} Compute the mean posterior diagonal mass $\bar{d}_{\text{exp}}$ and $\bar{d}_{\text{crowd}}$ for the two groups, and report their difference as a proxy for unsupervised separation quality.
\end{itemize}

\paragraph{Results}

\begin{figure}[H]
\centering
\includegraphics[width=0.8\linewidth]{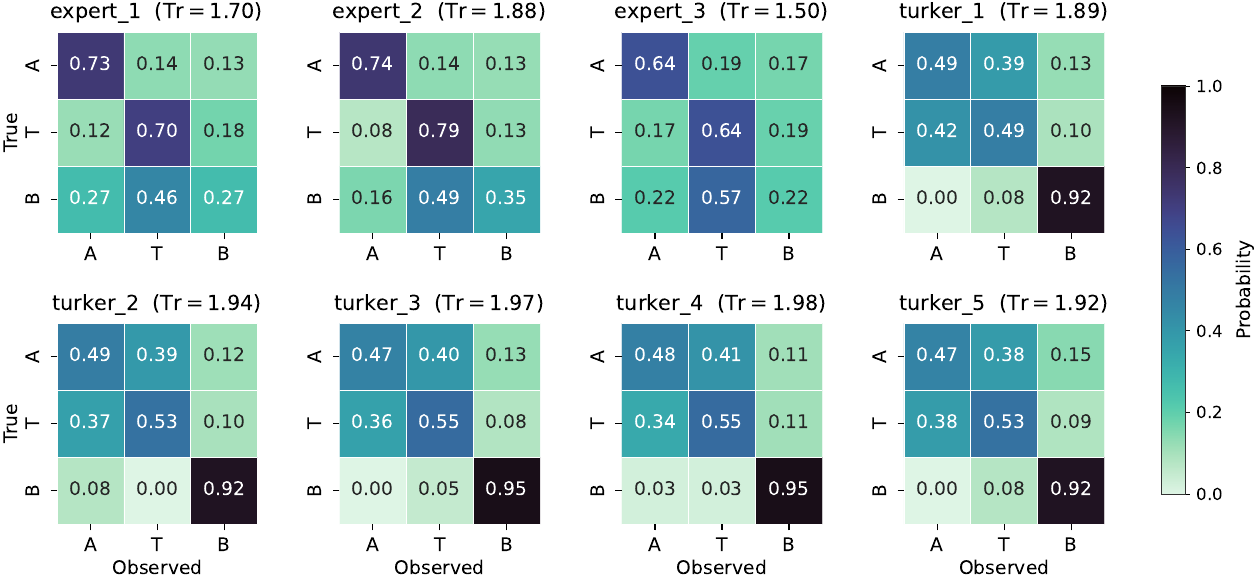}
\caption{Human confusion matrices (3 experts and 5 crowdworkers.)}
\label{appfig:human_conf_mat_8}
\end{figure}

Figure~\ref{appfig:human_conf_mat_8} displays the posterior confusion matrices $\hat{C}_r$ inferred on the full SummEval panel ($3$ experts, $5$ crowdworkers). Three structural patterns emerge. First, the experts (top row) exhibit \emph{diffuse} posteriors: diagonal entries are modest ($0.57$--$0.76$) with substantial off-diagonal leakage, particularly on the tie row (e.g., expert\_3 shows $0.20$ mass on $A$ and $0.46$ on $B$ when the true label is $T$). This reflects genuine quality disagreement among domain specialists. Second, the crowdworkers (bottom row) display \emph{concentrated}, nearly identical patterns: the tie-diagonal dominates ($0.44$--$0.54$) while $A$ and $B$ accuracies remain low ($0.34$--$0.42$), suggesting a mechanical ``when in doubt, call a tie'' heuristic. Notably, turker\_1 is an outlier with inverted $A$/$B$ preferences ($0.54$ on $B$ when true is $A$), which BayesJudge flags without any metadata.

\begin{figure}[H]
\centering
\includegraphics[width=0.8\linewidth]{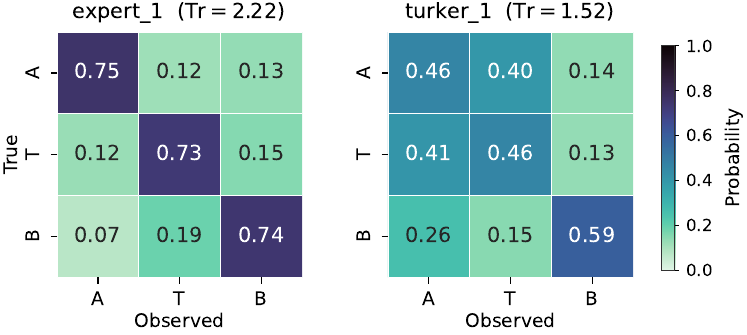}
\caption{Human confusion matrices (1 expert and 1 crowdworker.)}
\label{appfig:human_conf_mat_2}
\end{figure}

\begin{figure}[H]
\centering
\includegraphics[width=0.8\linewidth]{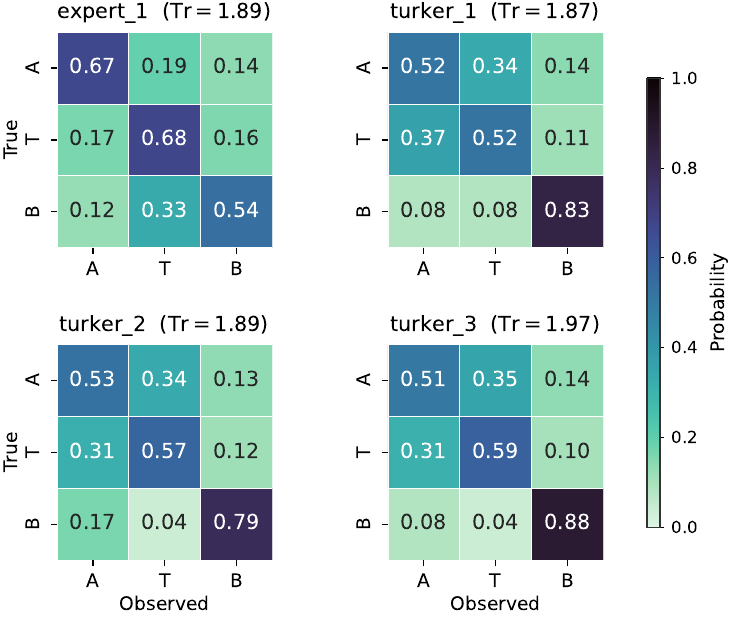}
\caption{Human confusion matrices (1 experts and 3 crowdworkers.)}
\label{appfig:human_conf_mat_4}
\end{figure}

\begin{figure}[H]
\centering
\includegraphics[width=0.8\linewidth]{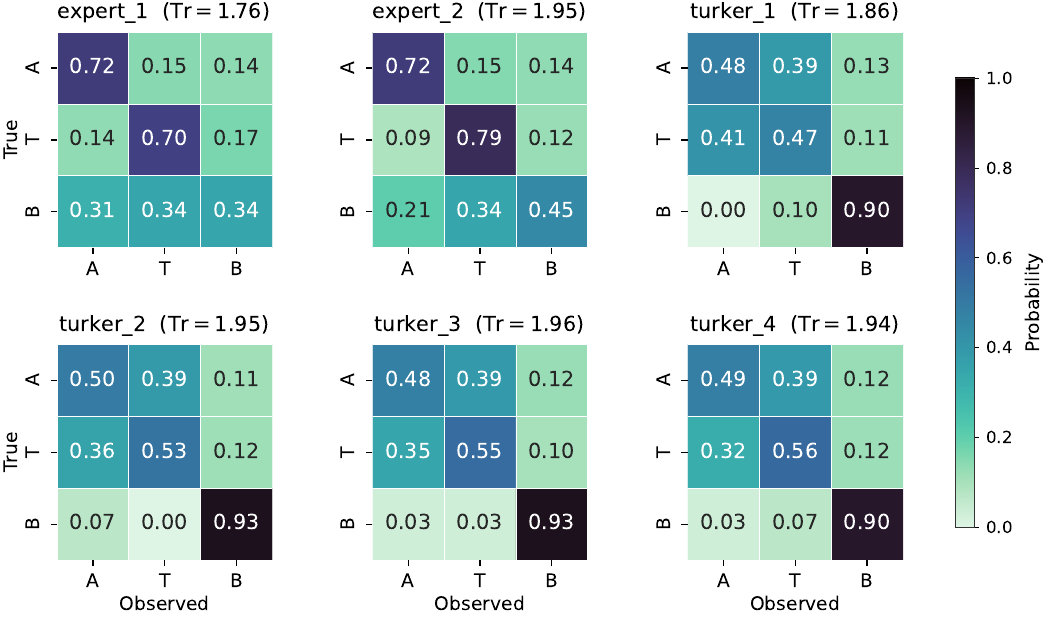}
\caption{Human confusion matrices (2 experts and 4 crowdworkers.)}
\label{appfig:human_conf_mat_6}
\end{figure}

Figures~\ref{appfig:human_conf_mat_2}--\ref{appfig:human_conf_mat_6} show the same inference under stratified subsamples. With only $1$ expert and $1$ crowdworker (Figure~\ref{appfig:human_conf_mat_2}), the two posterior matrices are barely distinguishable: the expert signature is contaminated by the sparse data, and the crowd pattern has not yet stabilized. At $1$ expert $+$ $3$ crowdworkers (Figure~\ref{appfig:human_conf_mat_4}), the crowdworkers' tie-concentration emerges, but the expert remains under-identified. Only at $2$ experts $+$ $4$ crowdworkers (Figure~\ref{appfig:human_conf_mat_6})---close to the true $3\!:\!5$ ratio---does the expert--crowd structural gap become visually salient: the experts retain diffuse off-diagonal mass while the crowdworkers lock onto the tie diagonal.

This progression validates two claims. First, BayesJudge's unsupervised rater typing is not an artifact of panel size; it requires a \emph{minimal per-type sample} ($\ge 2$ per group) to stabilize Dirichlet posteriors. Second, the expert--crowd divide is \emph{qualitative} rather than merely quantitative: it is not that experts are ``better'' (higher diagonal), but that they disagree in a \emph{structured} way (diffuse confusion), whereas crowdworkers agree on a \emph{simple} fallback heuristic (tie concentration). Majority vote conflates these qualitatively different noise patterns; BayesJudge separates them.

\subsection{Why Does Fusion Trade Crowd Accuracy for Expert Accuracy?}
\label{app:coupling}

The crowd-accuracy drop in Table~\ref{tab:heldout} might suggest that the LLM signal corrupts
crowdworker modeling at test time. We show instead that it is a training-time
effect arising from how the two channels are coupled.

\paragraph{Coupling mechanism.} Although the parameters are represented
separately, they are not inferred independently. The joint local objective is
\begin{equation}
\mathcal L_t(\eta) = \log p(\eta) + \sum_{r\in R_t}\log \Big(\textstyle\sum_k g_k(\eta)\,C_r[k, y_{tr}]\Big) + \textstyle\sum_j \log L^{\mathrm{swap}}_{tj}(\eta, \beta_j).
\end{equation}
Human labels enter through $C_r$, while the LLM swap likelihood acts directly
on the same $\eta$. Strong LLM evidence therefore shifts both $\hat\eta_t$ and
the latent row to which human observations are assigned when updating $C_r$:
the two channels are coupled through the shared latent state even though
their parameters are separate.

\paragraph{Verification.} We train the model on human labels only (yielding
$C_H$) and jointly with the LLM (yielding $C_J$), then predict held-out
raters under identical human-only evidence, varying only the confusion
matrix. With $C_H$, expert and crowd accuracies are $43.7\%$ and $68.9\%$;
replacing $C_H$ with $C_J$ alone reverses them to $67.0\%$ and $43.3\%$,
nearly matching the full joint model's $42.3\%$ crowd accuracy despite no LLM
signal being present at inference. The crowd prediction drop is therefore
attributable to confusion matrices being re-estimated during joint training:
fusion shifts the inferred consensus toward the expert-aligned notion of
quality, whereas in the human-only regime the crowd majority dominates the
consensus and $C_H$ consequently overestimates crowd reliability. On
SummEval's imbalanced panel (3 experts vs.\ 5 crowdworkers), this trade-off
is the price of refusing to let numerical superiority dictate the consensus.

\subsection{Multi-LLM Evaluator}
\label{app:multillm}
\begin{figure}[H]
\centering
\includegraphics[width=0.8\linewidth]{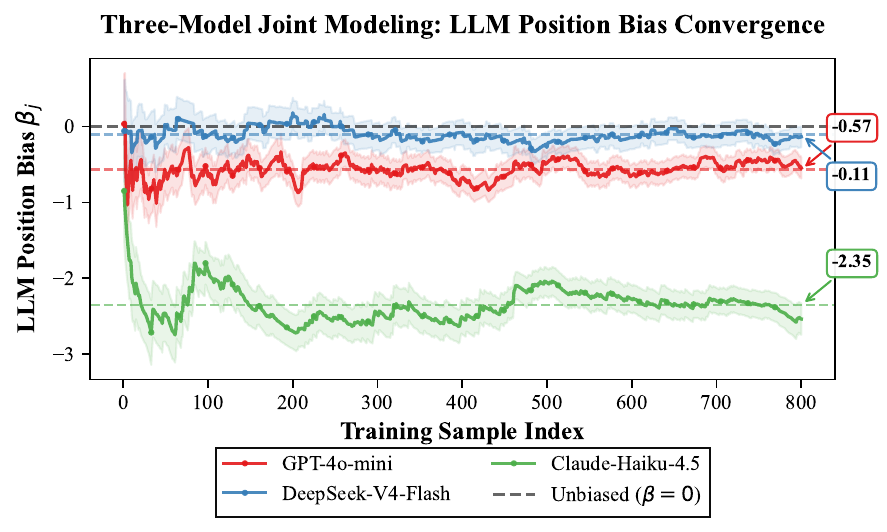}
\caption{Posterior mean trajectories of LLM position bias $\hat{\beta}_j$ for three jointly modeled judges on the SummEval stream. DeepSeek-V4-Flash (blue) converges near zero ($\hat{\beta}\approx -0.11$), indicating negligible position dependence. GPT-4o-mini (red) settles at $\hat{\beta}\approx -0.57$, reflecting a moderate post-position preference. Claude-Haiku-4.5 (green) exhibits a strong post-position bias ($\hat{\beta}\approx -2.35$). The clear separation and stable convergence of the three trajectories demonstrate that the paired swap likelihood successfully distinguishes and identifies heterogeneous judge-specific position effects.}
\label{appfig:3_model_beta}
\end{figure}

To test whether BayesJudge generalizes beyond a single model family, we jointly evaluate three LLM judges on the SummEval stream: GPT-4o-mini, Claude-Haiku-4.5, and DeepSeek-V4-Flash, all queried at temperature $0.0$ to isolate systematic position effects from sampling noise. Figure~\ref{appfig:3_model_beta} plots the online posterior mean $\hat{\beta}_j$ for each judge. The three models exhibit starkly heterogeneous bias profiles. DeepSeek-V4-Flash is effectively position-invariant, converging to $\hat{\beta} \approx -0.11$ with a credible band that includes zero, indicating that its swap judgments are driven primarily by intrinsic item quality. GPT-4o-mini displays a moderate post-position preference ($\hat{\beta} \approx -0.57$). Most strikingly, Claude-Haiku-4.5 exhibits a strong and persistent post-position bias ($\hat{\beta} \approx -2.35$), meaning its verdicts are heavily influenced by which candidate is presented second. The shaded posterior credible regions contract monotonically for all three judges, confirming that the paired swap protocol furnishes sufficient antisymmetric evidence to identify and calibrate family-specific distortions in streaming settings. This heterogeneity is precisely why joint modeling matters: when judges with divergent bias structures are fused, their opposing presentation effects mutually cancel, yielding a more reliable item posterior than any single judge could provide alone. Table \ref{tab:swap-deviation} shows the actual position bias ratios of the three models in the experimental results. It can be seen that BayesJudge's swap-likelihood design effectively identifies the corresponding position bias types.

\begin{table}[h]
\centering
\caption{Swap-level position bias statistics on SummEval. \textit{Biased samples} exclude ties and identical AB/BA responses. \textit{Inconsistent} denotes non-systematic deviations (e.g., non-AA/BB patterns). \textit{Within decisive} restricts to samples with unambiguous AA or BB preference.}
\label{tab:swap-deviation}
\small
\begin{tabular}{lccccccc}
\toprule
& \textbf{Biased} & \multicolumn{2}{c}{\textbf{Front (AA)}} & \multicolumn{2}{c}{\textbf{Back (BB)}} & \textbf{Inconsistent} & \textbf{Back \% within} \\
\textbf{Judge} & \textbf{samples} & n & \% & n & \% & n (\%) & \textbf{decisive} \\
\midrule
GPT-4o-mini & 329 & 27 & 8.21\% & 273 & 82.98\% & 29 (8.81\%) & 91.00\% \\
DeepSeek-V4-Flash & 142 & 38 & 26.76\% & 68 & 47.89\% & 36 (25.35\%) & 64.15\% \\
Claude-Haiku-4.5 & 908 & 1 & 0.11\% & 794 & 87.44\% & 113 (12.44\%) & 99.87\% \\
\bottomrule
\end{tabular}
\end{table}

\section{MT-Bench Evaluation}
\label{app:mtbench}

To test whether BayesJudge's effectiveness is tied to a particular dataset or domain, we evaluate on MT-Bench, a multi-turn dialogue benchmark. We use the samples that contain exactly 2 human annotations, yielding 303 comparisons, split 8:2 into training and test sets. We follow the identical protocol as the SummEval experiments: LOHO evaluation, the same priors, and GPT-4o-mini as the LLM judge under AB/BA swaps.

\begin{table}[h]
  \caption{Held-out prediction on MT-Bench (303 samples, 2 human annotators per item). With only two annotations per item, LOHO leaves a single human label as evidence without rater-specific parameters, BTD's prediction reduces to majority vote in this regime.
}
  \label{tab:mtbench}
  \centering
  \small
  \begin{tabular}{lcccc}
    \toprule
    Method & Acc $\uparrow$ & NLL $\downarrow$ & Brier $\downarrow$ & ECE $\downarrow$ \\
    \midrule
    Majority-Vote & 0.656 & 9.512 & 0.689 & 0.344 \\
    BTD & 0.656 & 9.512 & 0.689 & 0.344 \\
    MACE & 0.656 & 0.855 & 0.494 & \textbf{0.069} \\
    GLAD & 0.672 & 0.885 & 0.489 & 0.142 \\
    \midrule
    \textbf{BayesJudge (ours)} & \textbf{0.713} & \textbf{0.809} & \textbf{0.468} & 0.134 \\
    \bottomrule
  \end{tabular}
\end{table}

BayesJudge achieves the best accuracy (71.3\%, $+4.1$ points over the strongest baseline), NLL, and Brier score on this dataset, and its calibration advantage over non-probabilistic baselines (NLL $=9.5$ for voting-based methods) is consistent with the SummEval findings. The effectiveness of BayesJudge is therefore not tied to a particular dataset or domain, although both real-world evaluations remain within pairwise evaluation of open-ended generation.

\clearpage

\section*{NeurIPS Paper Checklist}

\begin{enumerate}

\item {\bf Claims}
    \item[] Question: Do the main claims made in the abstract and introduction accurately reflect the paper's contributions and scope?
    \item[] Answer: \answerYes{}.
    \item[] Justification: See the abstract and Section \ref{sec:intro}.

\item {\bf Limitations}
    \item[] Question: Does the paper discuss the limitations of the work performed by the authors?
    \item[] Answer: \answerYes{}.
    \item[] Justification: See Section \ref{sec:discussion}.

\item {\bf Theory assumptions and proofs}
    \item[] Question: For each theoretical result, does the paper provide the full set of assumptions and a complete (and correct) proof?
    \item[] Answer: \answerYes{}.
    \item[] Justification: See Section \ref{sec:theory} and Appendix \ref{app:proof}.

    \item {\bf Experimental result reproducibility}
    \item[] Question: Does the paper fully disclose all the information needed to reproduce the main experimental results of the paper to the extent that it affects the main claims and/or conclusions of the paper (regardless of whether the code and data are provided or not)?
    \item[] Answer: \answerYes{}.
    \item[] Justification: See the abstract for the code url.

\item {\bf Open access to data and code}
    \item[] Question: Does the paper provide open access to the data and code, with sufficient instructions to faithfully reproduce the main experimental results, as described in supplemental material?
    \item[] Answer: \answerYes{}.
    \item[] Justification: See the abstract for the code url.

\item {\bf Experimental setting/details}
    \item[] Question: Does the paper specify all the training and test details (e.g., data splits, hyperparameters, how they were chosen, type of optimizer) necessary to understand the results?
    \item[] Answer: \answerYes{}.
    \item[] Justification: See the Section \ref{sec:experiments} and Appendix \ref{app:sensitive}.

\item {\bf Experiment statistical significance}
    \item[] Question: Does the paper report error bars suitably and correctly defined or other appropriate information about the statistical significance of the experiments?
    \item[] Answer: \answerYes{}.
    \item[] Justification: See Section \ref{sec:experiments}.

\item {\bf Experiments compute resources}
    \item[] Question: For each experiment, does the paper provide sufficient information on the computer resources (type of compute workers, memory, time of execution) needed to reproduce the experiments?
    \item[] Answer: \answerYes{}.
    \item[] Justification: See Appendix \ref{app:sensitive}.
    
\item {\bf Code of ethics}
    \item[] Question: Does the research conducted in the paper conform, in every respect, with the NeurIPS Code of Ethics \url{https://neurips.cc/public/EthicsGuidelines}?
    \item[] Answer: \answerYes{}.
    \item[] Justification: See Section \ref{sec:discussion}.

\item {\bf Broader impacts}
    \item[] Question: Does the paper discuss both potential positive societal impacts and negative societal impacts of the work performed?
    \item[] Answer: \answerYes{}.
    \item[] Justification: Our work may affect the human-LLM evaluation process.
    
\item {\bf Safeguards}
    \item[] Question: Does the paper describe safeguards that have been put in place for responsible release of data or models that have a high risk for misuse (e.g., pre-trained language models, image generators, or scraped datasets)?
    \item[] Answer: \answerNA{}.
    \item[] Justification: This paper poses no such risks.

\item {\bf Licenses for existing assets}
    \item[] Question: Are the creators or original owners of assets (e.g., code, data, models), used in the paper, properly credited and are the license and terms of use explicitly mentioned and properly respected?
    \item[] Answer: \answerYes{}.
    \item[] Justification: See Section \ref{sec:experiments}.

\item {\bf New assets}
    \item[] Question: Are new assets introduced in the paper well documented and is the documentation provided alongside the assets?
    \item[] Answer: \answerYes{}.
    \item[] Justification: See abstract for the code url.

\item {\bf Crowdsourcing and research with human subjects}
    \item[] Question: For crowdsourcing experiments and research with human subjects, does the paper include the full text of instructions given to participants and screenshots, if applicable, as well as details about compensation (if any)? 
    \item[] Answer: \answerNA{}.
    \item[] Justification: The paper does not involve crowdsourcing nor research with human subjects.

\item {\bf Institutional review board (IRB) approvals or equivalent for research with human subjects}
    \item[] Question: Does the paper describe potential risks incurred by study participants, whether such risks were disclosed to the subjects, and whether Institutional Review Board (IRB) approvals (or an equivalent approval/review based on the requirements of your country or institution) were obtained?
    \item[] Answer: \answerNA{}.
    \item[] Justification: The paper does not involve crowdsourcing nor research with human subjects.

\item {\bf Declaration of LLM usage}
    \item[] Question: Does the paper describe the usage of LLMs if it is an important, original, or non-standard component of the core methods in this research? Note that if the LLM is used only for writing, editing, or formatting purposes and does \emph{not} impact the core methodology, scientific rigor, or originality of the research, declaration is not required.
    \item[] Answer: \answerYes{}.
    \item[] Justification: We used LLMs for evaluation tasks in our experiments.

\end{enumerate}

\end{document}